\documentclass[numberedappendix]{emulateapj}
\usepackage{aas_macros}
\usepackage{float}
\usepackage{amsmath,amssymb}
\usepackage{natbib}    
\usepackage{hyperref} 
\usepackage{graphicx} 
\usepackage{color}
\usepackage{verbatim}

\newcommand{\proptosim}{\mathrel{\vcenter{
  \offinterlineskip\halign{\hfil$##$\cr
    \propto\cr\noalign{\kern2pt}\sim\cr\noalign{\kern-2pt}}}}}

\renewcommand{\b}[1]{\boldsymbol{#1}}
\newcommand{\unit}[1]{{\rm #1}}

\newcommand{\method}{Wang 2017, in preparation}
\newcommand{\au}{\,\mbox{\sc au}}
\newcommand{\cm}{\unit{cm}}

\newcommand{\g}{\unit{g}}
\newcommand{\G}{\unit{G}}
\newcommand{\K}{\unit{K}}

\newcommand{\kms}{\unit{km\ s^{-1}}}

\renewcommand{\micron}{\unit{\mu m}}

\newcommand{\erg}{\unit{erg}}
\newcommand{\eV}{\unit{eV}}
\newcommand{\keV}{\unit{keV}}
\newcommand{\s}{\unit{s}}
\newcommand{\yr}{\unit{yr}}
\newcommand{\ang}{\ensuremath{\mathrm{\AA}}}

\newcommand{\cs}{c_{\rm s}}
\newcommand{\dash}{\mbox{--}}
\newcommand{\lya}{\text{Ly}\ensuremath{\alpha}}
\newcommand{\kb}{k_\mathrm{B}}
\newcommand{\sigb}{\sigma_\mathrm{B}}
\renewcommand{\d}{\mathrm{d}}

\newcommand{\Gr}{\ensuremath{\mathrm{Gr}}}

\newcommand{\eff}{\mathrm{eff}}
\newcommand{\diss}{\mathrm{diss}}
\newcommand{\wind}{\mathrm{wind}}
\newcommand{\dust}{\mathrm{dust}}
\newcommand{\disk}{\mathrm{disk}}
\newcommand{\de}{\mathrm{de}}   
\renewcommand{\sp}{\mathrm{sp}}   
\newcommand{\ah}{\mathrm{ah}}   
\newcommand*\chem[1]{\ensuremath{\mathrm{#1}}}
\newcommand{\pos}[1]{\ensuremath{\mathrm{#1}^+}}
\renewcommand{\neg}[1]{\ensuremath{\mathrm{#1}^-}}
\newcommand{\ext}[1]{\ensuremath{\mathrm{#1}^*}}
\begin{document}

\title{Hydrodynamic Photoevaporation of Protoplanetary Disks
with Consistent Thermochemistry}

\author{Lile Wang$^1$ and Jeremy J. Goodman$^1$}

\footnotetext[1]{Princeton University Observatory,
  Princeton, NJ 08544}

\begin{abstract}
  Photoevaporation is an important dispersal mechanism for
  protoplanetary disks.  We conduct hydrodynamic simulations
  coupled with ray-tracing radiative transfer and consistent
  thermochemistry to study photoevaporative winds driven by
  ultraviolet and X-ray radiation from the host star.  Most
  models have a three-layer structure: a cold midplane, warm
  intermediate layer, and hot wind, the last having typical
  speeds $\sim 30~\kms$ and mass-loss rates
  $\sim 10^{-9}~M_\odot~\yr^{-1}$ when driven
  primarily by ionizing UV radiation. Observable molecules
  including \chem{CO}, \chem{OH} and \chem{H_2O} re-form in
  the intermediate layer and survive at relatively high wind
  temperatures due to reactions being out of equilibrium.
  Mass-loss rates are sensitive to the intensity of
  radiation in energy bands that interact directly with
  hydrogen.  Comparison with previous works shows that mass
  loss rates are also sensitive to the treatment of both the
  hydrodynamics and the thermochemistry.  Divergent results
  concerning the efficiency of X-ray photoevaporation are
  traced in part to differing assumptions about dust and
  other coolants.
\end{abstract}


\section{Introduction}
\label{sec:intro}

Protostellar/protoplanetary disks (hereafter PPDs)
surrounding low-mass T~Tauri stars are the birthplaces of
planets and have typical lifetimes $\sim 10^6-10^7\ \yr$
lifespan \citep[e.g.][] {1995Natur.373..494Z,
  2001ApJ...553L.153H}.  Along with accretion onto the star,
sequestration of mass in planets, and perhaps
magnetized disk winds, photoevaporation by hard photons
likely contributes to the dispersal of PPDs
\citep{1994ApJ...428..654H}.

Hard photons in different energy bands experience different
microscopic physics and have differing effects on
PPDs. Following \citet{2009ApJ...690.1539G}, we use the term
``far-UV (FUV)'' for photon energies
$6~\eV < h\nu < 13.6~\eV$, ``extreme-UV (EUV)'' for
$13.6~\eV < h\nu < 0.1~\keV$, and ``X-ray'' for
$h\nu > 0.1~\keV$. While EUV may be blocked by the wind from
the disk surface \citep[e.g.][]{2005MNRAS.358..283A},
FUV and X-ray radiation are more penetrating.  All of these
heat, dissociate, or ionize the gas via a plethora of
mechanisms. In order to model photoevaporation of PPDs,
therefore, one is required to take the richness of the
microphysics into account, as well as its interaction with
the hydrodynamics.

Evolving a hydrodynamic system coupled with thermochemistry
to (quasi-) steady state could be prohibitively expensive if
a large chemical reaction network were included.  Past
work on PPD photoevaporation has compromised (at least) one
of the two aspects: hydrodynamics or thermochemistry.
\citet{2006MNRAS.369..216A, 2006MNRAS.369..229A}
modeled EUV photoevaporation in hydrodynamic simulations
with minimum thermochemistry.  On
the other hand, calculations with detailed thermochemistry
usually adopt semi-analytic prescriptions for the wind
mass-loss rate rather than simulate multidimensional
hydrodynamics e.g. \citet{2008ApJ...683..287G,
  2009ApJ...690.1539G} (hereafter GH08, GH09).  Some recent
works conduct hydrodynamic simulations with interpolation
tables for gas temperature
drawn from hydrostatic
scenarios \citep[e.g.][]{2010MNRAS.401.1415O}.  More
recently, \citet{2012MNRAS.420..562H},
\citet{2016MNRAS.463.3616H}, and \citet{2017MNRAS.468L.108H}
have coupled hydrodynamics and thermochemistry in
simulations of externally irradiated disks and
pre-stellar cores; their code is three-dimensional, but
their applications have been confined mostly to simplified
geometries (spherical or cylindrical) for easier comparison
to semi-analytic work.

This work focuses on a consistent combination of
hydrodynamic simulation with a moderate-scale chemical
network (24 species, $\sim 10^2$ reactions). We include the
species and reactions that are relevant to photoevaporation,
especially heating and cooling mechanisms.  Full
hydrodynamic simulations are carried out in 2.5-dimensions
(axisymmetry), coupled with radiation, thermodynamics, and
chemistry, by solving time-dependent differential equations
in every zone throughout the simulation domain. Compared to
simulations with interpolation tables for thermochemistry,
this approach is able to deal with non-equilibrium
processes, as when some chemical and hydrodynamic timescales
are comparable.  The long-term goal of our exploration is to
predict observables, especially emission and absorption-line
profiles and strengths of important atomic and molecular
species, thereby constraining our wind models and the
parameters that go into them (e.g. abundances, dust
properties, EUV luminosities). We aim eventually to
incorporate MHD processes, and expect that the combination
of photoevaporative and magnetic effects will lead to higher
mass-loss rates than each process acting alone.  The
hydrodynamic simulations presented here are first steps
toward these goals.

This paper is structured as follows. In \S\ref{sec:methods},
we briefly summarize our numerical methods and physical
approximations. Additional details concerning our treatment
of thermochemical processes are given in the Appendices.
\S\ref{sec:fiducial-model} introduces the parameter choices
underlying our fiducial model. \S\ref{sec:results} presents
the main results of our calculations for this model, and for
several other models that differ from the fiducial one in
one or more parameters, with the goal of exploring the
effects of these parameters on gross properties of the flow,
especially the mass-loss rate.  In \S\ref{sec:discussions},
we discuss the role that different bands of radiation play,
and also compare and contrast our results with those of
\citet{2009ApJ...690.1539G} and \cite{2010MNRAS.401.1415O}.
\S\ref{sec:conclusion} concludes and summarizes the paper.

\section{Methods}
\label{sec:methods}

This section summarizes our methods.  The computational
scheme for hydrodynamics is first described, followed by our
methods for radiative transfer and thermochemistry.

\subsection{Hydrodynamics}
\label{sec:numerical-method}

Our modeling of PPD photoevaporation systems involves full
hydrodynamic calculations. We use the grid-based,
general-purpose, astrophysical code \verb|Athena++|
(\citealt{2016ApJS..225...22W}; Stone et al., in
preparation) in spherical coordinates $(r,\theta,\phi)$ but
neglect all dependence on $\phi$: our simulations are
axisymmetric.  Magnetic fields are neglected in the present
work, although \verb|Athena++| is fully capable of MHD
(indeed optimized for it).  We use the HLLC Riemann solver
and van Leer reconstruction with improved order of accuracy
using the revised slope limiter
\citep[see][]{2014JCoPh.270..784M}.  Consistent Multi-fluid
Advection (CMA) is used to ensure strict conservation of
chemical elements and species \citep[e.g.][]
{2010MNRAS.404....2G}.

\subsection{Radiative transfer}
\label{sec:ray-tracing}

Absorption processes dominate scattering for most of the
radiation that we consider: FUV, EUV, and $1~\mathrm{keV}$
X-rays \citep{Draine_book, Verner+etal1996}.  An exception
would be $\lya$ photons, which may dominate the FUV
luminosity, and whose scattering into nonradial directions
helps them to penetrate more deeply into the disk
\citep[e.g.][]{2011ApJ...739...78B}.  We find, however, that
unscattered soft FUV photons penetrate the intermediate
layer anyway, and more deeply than $\lya$.  Like $\lya$,
these photons dissociate \chem{H_2O} and \chem{OH}, which
can be important coolants, but not \chem{H_2} or
\chem{CO}\citep[e.g.][]{1978ApJ...224..841S}.  The
scattering of harder X-rays can be important for ionization
and hence magnetic coupling of the upper layers of the disk
\citep[e.g.]{Igea+Glassgold1999,Bai+Goodman2009}, but we are
neglecting magnetic fields here.

Therefore, in this paper, scattering is neglected, and
radiative transfer consists only of radial ray tracing, the
sources of all hard photons being assumed to lie at the
origin ($r=0$).  This is facilitated by our choice of
spherical coordinates, although our algorithm can trace rays
in nonradial directions also (\method).

One ray is assigned to each radial column.  Its luminosity
is adjusted as it propagates through each cell according to
the photoreactions within that cell.  Some cells can be
individually optically thick. Hence for photochemistry, we
adopt as the effective flux at photon frequency $\nu$,
\begin{equation}
  \label{eq:eff-flux}
  F_\eff(\nu) = F_0(\nu)
  \left\{ \dfrac{1 - \exp[-\delta l / \lambda(\nu)]}
    {\delta l / \lambda(\nu)} \right\}\ ,
\end{equation}
where $F_0$ is the flux impinging on the inner face of the
current cell, $\lambda(\nu)$ is the local absorption mean
free path of photons at frequency $\nu$, and $\delta l$ is
the chord length of the ray across the cell. (For radial ray
tracing, $\delta l$ is simply the radial width of the cell.)
Eq.~\eqref{eq:eff-flux} yields $F_\eff\rightarrow F_0$ as
$(\delta l / \lambda) \rightarrow 0$.

\subsection{Chemistry and Thermodynamics}
\label{sec:chem-thermo}

In each cell, a coupled set of ordinary differential
equations (ODEs) is solved to update the abundances of all
$\mathcal{N}$ chemical species $\{n^i\}$ and internal energy
density $\epsilon$.  These equations read, nominally,
\begin{equation}
  \label{eq:ode-chem-thermo}
  \begin{split}
    \dfrac{\d n^i}{\d t} & = \mathcal{A}^i_{\;jk} n^j n^k +
    \mathcal{B}^i_{\;j} n^j\ ; \\
    \dfrac{\d \epsilon}{\d t} & = \Gamma - \Lambda\ ;
  \end{split}
\end{equation}
in which the terms involving $\{\mathcal{A}^i_{\;jk}\}$
describe two-body reactions, while those in
$\{\mathcal{B}^i_{\;j}\}$ represent photoionization and
photodissociation.  $\Gamma$ and $\Lambda$ are the heating
and cooling rates per unit volume,
respectively. $\{\mathcal{A}^i_{\;jk}\}$, $\Gamma$ and
$\Lambda$ are usually functions of temperature $T$.  The
thermal energy density $\epsilon = c_V(\{n^i\}) T$, where
$c_V$ is the heat capacity of the gas at constant volume.
(Thermochemistry and hydrodynamics are solved in separate
substeps, whence we use $c_V$ instead of $c_P$ here.)  The
ODEs \eqref{eq:ode-chem-thermo} are solved in conjunction
with the hydrodynamics by operator splitting.  That is, they
are advanced one time step after each hydrodynamic step,
which has included advection of the chemical species, while
holding the masses of all elements fixed within each cell.
Photoreactions are included using the radiative fluxes
computed as described in \S\ref{sec:ray-tracing}.  The
updated internal energy $\epsilon$ and number densities
$\{n^i\}$ of all species are then used to initialize the
next hydrodynamic step.

The ODEs \eqref{eq:ode-chem-thermo} are usually stiff and
hence numerically difficult. We use a standard ODE solver
with adaptive implicit modules, \verb|CVODE|
\citep[see][]{hindmarsh2005sundials}.  The solution of these
equations dominates our total computation time, typically by
a factor $\gtrsim 10$ compared to the hydrodynamics.
Nonetheless, this brute-force approach rewards us by being
able to deal with non-equilibrium conditions, as will be
discussed later in this paper.

Guided by GH08, GH09, and our own numerical experiments, we
adopt 24 species that are most relevant to heating and
cooling processes involved in PPD photoevaporation: \neg{e}
(free electrons), \pos{H}, H, \chem{H_2}, \ext{\chem{H_2}}
(using the $v=6$ vibrational state as a proxy for \chem{H_2}
in all excited states, see Appendix~\ref{sec:fuv-h2} and
TH85), He, \pos{He}, O, \pos{O}, \ext{O} (the $^1D$ state of
atomic oxygen as a proxy for all neutral excited states, see
Appendix \ref{sec:fuv-oh}), OH, \chem{H_2O}, C, \pos{C}, CO,
S, \pos{S}, Si, \pos{Si}, Fe, \pos{Fe}, Gr, \pos{Gr},
\neg{Gr}. Here Gr and Gr$^\pm$ denote neutral and
singly-charged dust grains, respectively.

We extract the reactions involving these species from the
\verb|UMIST| astrochemistry database
\citep{UMIST2013}. However, the interstellar radiation
fields and matter densities to which the standard
\verb|UMIST| database is usually applied are rather
different from those of PPDs.  We therefore exclude all
reactions involving photons and dust grains in the
\verb|UMIST| library; instead, we evaluate those reaction
rates separately.

Photoionization and photodissociation are critical
mechanisms that affect photoevaporation. At each photon
energy, the ionization cross section of each atomic species
is evaluated using the data in \citet{Verner+Yakovlev1995,
  Verner+etal1996}. For molecular species that can react
with FUV photons, namely \chem{H_2}, CO, OH, and \chem{H_2O}
here, we adopt the FUV-induced photochemcial reaction rate
based on \citet[hereafter TH85]{1985ApJ...291..722T} for
\chem{H_2}, \citet{2009A&A...503..323V} for CO (note that
this photodissociation cross section is $\sim 10\times$ the
value in TH85), and \citet[hereafter
AGN14]{2014ApJ...786..135A} for \chem{H_2O} and OH. The
photochemical processes related to \chem{H_2}, C and CO may
be subject to considerable self-shielding and
cross-shielding. Using the radial column density data that
are obtained by integrating along radial rays, we evaluate
the impact of the self-/cross-shielding by adopting the
analytic formulae in \citealt{2009A&A...503..323V} (for CO)
and TH85 (for C), and \citet{1996ApJ...468..269D} (for
\chem{H_2}). It is worth noting that the FUV-induced
processes in parallel with photodissociation of \chem{H_2},
\chem{H_2O} and OH, e.g. FUV pumping of \chem{H_2} onto its
excited states and its subsequent effects, can have
considerable thermodynamic effects. We refer the reader to
Appendices \ref{sec:fuv-h2} and \ref{sec:fuv-oh} for
detailed discussion.

Heating and cooling processes are directly
associated with chemical reactions. While the amount of
energy deposited into and removed from the gas by
photoionization and recombination can be estimated
straightforwardly \citep[see also][eqs. 27.3, 27.23]
{Draine_book}, the thermodynamic effects of other chemical
reactions need elaboration, which is provided in
Appendices~\ref{sec:fuv-h2} through \ref{sec:dust-pah}.
There are other radiative mechanisms that remove energy from
the gas, especially collisionally pumped ro-vibrational
transitions of molecules, and fine-structure transitions of
atoms. We briefly summarize those mechanisms and our method
for evaluating them in Appendix \ref{sec:mol-atom-cooling}.

Dust grains are usually crucial in PPD
photoevaporation. Following the arguments in GH08, as well
as \citet{2006A&A...459..545G, 2007A&A...476..279G,
  2011ApJ...727....2P}, we suggest that polycyclic aromatic
hydrocarbons (PAHs) overwhelm dust grains of other sizes in
terms of the following effects, thanks to their dominant
contribution to total dust surface area:
photoelectric heating of gas, dust-gas collisional energy
transfer, recombination with free electrons, dust-assisted
molecular hydrogen formation, and neutralization of positive
ions. We include the processes listed above as outlined in
Appendix \ref{sec:dust-pah}.

\section{Choice of Fiducial model}
\label{sec:fiducial-model}

This section presents the setup of our fiducial model, whose
main properties are listed in Table
\ref{table:fiducial-model}.  Other models, each differing
from the fiducial in one parameter, are described in
\S\ref{sec:various-model}.

The simulation domain is axisymmetric, extending from
$2~\au$ to $100~\au$ in radius ($r$) and $0$ to $\pi/2$ in
colatitude ($\theta$).  All models are presumed to be
symmetric about the equatorial plane, so that, for example,
quoted mass-loss rates include outflows at $\theta>\pi/2$.
All dependence on the azimuthal coordinate ($\phi$) is
ignored. Outflow boundary conditions with a radial flow
limiter (which inhibits radial inflow) are imposed at
$r=2~\au$ and $r=100~\au$, and reflecting boundary
conditions at $\theta=0$ and $\theta = \pi/2$.  Our standard
resolution is $256$ radial by $128$ latitudinal zones, the
radial zones being logarithmically spaced, and the
latitudinal zones equally spaced.

The gravitational field is that of a $1~M_\odot$ star
located at the origin.  The disk, whose self-gravity is
neglected, is initialized in hydrostatic and centrifugal
balance, except for slight imbalances due to numerical
discretization.  The disk density and temperature profile
follow the steady state solution in
\citet{2013MNRAS.435.2610N}, in which we set the midplane
density as $n=10^{10}~\cm^{-3}$ and temperature $T=20~\K$ at
$r=10^2~\au$, with radial power index being $(-2.25)$ for
density and $(-0.5)$ for temperature--this profile yields a
disk mass $\approx 0.03 M_\odot$ within $100~\au$.  The
density and temperature profiles roughly agree with GH09,
but the latter are not quite hydrostatic.

All radiation emanates from the origin of spherical polar
coordinates.  Our simulation domain does not cover the
origin, and the rays are not attenuated before they reach
the inner boundary. The source is isotropic, but those rays
that reach the midplane region at the inner boundary are
discarded (we also test not discarding those rays, finding
negligible differences in the dissociation layer and in the
wind). Each ray has four discrete energy bins, representing
four important bands of photon energy: $h\nu = 7~\eV$ for
FUV photons that do not interact appreciably with hydrogen
molecules (``soft FUV'' hereafter), $12~\eV$ for
Lyman-Werner (``LW'' for short) band photons, $25~\eV$ for
EUV photons, and $1~\keV$ for X-ray photons. $\lya$ photons
are neglected, as discussed above.  The number of photons
radiated in each energy bin per unit time follows the
luminosity model described in GH08 and GH09: (1) a $9000~\K$
black body spectral profile for FUV
($6~\eV<h\nu < 13.6~\eV$) with total luminosity
$L_\mathrm{FUV} = 10^{31.7}~\erg~\s^{-1}$; (2) an additional
EUV-photon emission rate \footnote{GH08 and GH09 assumed
  different EUV luminosities for their fiducial models. Here
  we adopt the value specified in GH09.}
$\Phi_\mathrm{EUV} = 10^{40.7}~\s^{-1}$; (3) and an X-ray
luminosity $L_X=10^{30.4}~\erg~\s^{-1}$.

The initial elemental abundances are determined by the
values in Table \ref{table:fiducial-model} (wherein
$n_{\chem{H}}$ is the number density of hydrogen nuclei).
These choices generally follow a subset of those in GH08,
with the additional assumption that elements appear in
chemical compounds if possible.  These initial abundances
are uniform throughout the simulation domain.

Our assumptions about the dust turn out to be important for
our results.  GH08 and GH09 treated two populations of
grains: (i) an MRN-type power-law distribution with a
minimum grain radius of $50~\ang$, maximum of
$20~\micron$, and a dust-to-gas ratio by mass of
$10^{-2}$; and (ii) PAH grains with abundance
$8.4\times10^{-8}$ per hydrogen nucleus.  The first
population has a total geometrical cross section of
$2\times10^{-22}~\cm^2$ per hydrogen nucleus
($\sigma_{\rm dust}/\chem{H}$).  The authors do not state
the radius of their PAH grains explicitly, but they refer to
\citet{2001ApJ...554..778L}, and we interpret this to mean
that their PAHs can be approximated by spheres of radius
$6~\ang$.  It would follow that the contribution of their
PAHs to the cross section is
$\sigma_{\rm dust}/\chem{H}\approx 9.5\times 10^{-22}$, i.e.
several times larger than that of their MRN population,
although the contribution to the dust-to-gas mass ratio is
only $\sim10^{-4}$.  As noted above, \ref{sec:dust-pah}, the
principal effects of dust, especially heating and absorption
of radiation, are expected to be dominated by the smallest
grains---PAHs.

For simplicity, we prefer to work with a single-sized grain
population.  We therefore neglect MRN grains and take the
approximate relative abundance for our PAH-like grain
species (Gr) as $10^{-7}$ per hydrogen atom, slightly
greater than that of GH08, and a PAH radius of $5~\ang$,
i.e. approximately $60$ carbon atoms per PAH: see
\citealt{2001ApJS..134..263W}). The dust-to-gas mass ratio
is then $0.7\times 10^{-4}$, and
$\sigma_{\rm dust}/\chem{H}=8\times10^{22}~\cm^2$.

Although variable dust abundance is fully allowed by our
code, for the sake of simplicity we set the relative
abundance of Gr to be uniform
and assume that the dust comoves with the gas.

We run the simulation for $1.2\times 10^4~\yr$ with
microphysics enabled but the central radiation sources
turned off until the disk structure is fully numerically
relaxed, and the temperature profile converges to that set
by the artificial heating profile
(\S\ref{sec:dust-gas-heating}, which is sufficiently close
to the initial profile. The chemical abundances do not
change during this relaxation process except by passive
advection.  We confirm after this process that the disk is
indeed in hydrodynamic equilibrium and has no outflow.
Then, at $t= 3.6\times 10^3~\yr$, irradiation is turned on
and remains on for the rest of simulation; this lasts
$\gtrsim 500~\yr$, sufficiently long compared to the radial
flow timescale $\tau\sim (100~\au)/(30~\kms)\approx 16~\yr$
so as to reach an approximate quasi-steady state. On
Princeton University's local computer cluster
\verb|perseus|, $500~\yr$ of simulated time takes
$\sim 100~\mathrm{\ hrs}$ of wall-clock time on 128 CPUs.
About 95 per cent of the time is consumed by the thermochemical
calculations for the fiducial model, the hydrodynamic and
ray-tracing steps being relatively quick.

\begin{deluxetable}{lr}
  \tablecolumns{2} 
  \tabletypesize{\scriptsize}
  \tablewidth{0pt}
  \tablecaption{Properties of the fiducial model}   
  \tablehead{
    \colhead{Item} &
    \colhead{Value}
  }
  \startdata
  Radial domain & $2~\au \le r \le \ 100~\au$\\
  Latitudinal domain & $0\le\theta\le\pi/2$ \\
  Resolution & $N_{\log r} = 256$, $N_\theta= 128$ \\
  \\
  Stellar mass & $1.0~M_\odot$ \\
  \\
  $M_\disk$ & $0.03~M_\odot$ \\[2pt]
  Mid-plane density &
  $10^{10}(R/100~\au)^{-2.25}~\cm^{-3}$ \\[2pt]
  Mid-plane temperature &
  $20(R/100~\au)^{-0.5}~\K$ \\
  \\
  Luminosities [photon~$\s^{-1}$] & \\[5pt]
  $7~\eV$ (``soft'' FUV)  & $4.5\times 10^{42}$ \\
  $12~\eV$ (LW) & $1.6\times 10^{40}$ \\ 
  $25~\eV$ (EUV) & $5.0\times 10^{40}$ \\
  $1~\keV$ (X-ray) & $1.6\times 10^{39}$ \\
  \\
  Initial abundances [$n_{\chem{X}}/n_{\chem{H}}$] & \\[5pt]
  \chem{H_2} & 0.5\\
  He & 0.1\\
  \chem{H_2O} & $1.8 \times 10^{-4}$\\
  CO & $1.4 \times 10^{-4}$\\
  S  & $2.8 \times 10^{-5}$\\
  Si & $1.7 \times 10^{-6}$\\
  Fe & $1.7 \times 10^{-7}$\\
  Gr & $1.0 \times 10^{-7}$ \\
  \\
  Dust/PAH properties & \\
  $r_\dust$ & $5~\ang$ \\
  $\rho_\dust$ & $2.25~\g~\cm^{-3}$ \\
  $m_\dust/m_\mathrm{gas}$ & $7\times 10^{-5}$ \\
  $\sigma_\dust/\chem{H}$ & $8\times 10^{22}~\cm^2$ \\
  \enddata
  \label{table:fiducial-model}
\end{deluxetable}

We also calculate several models that differ from the
fiducial in one or more parameters, as described in
\S\ref{sec:various-model}.

\section{Results}
\label{sec:results}

In this section, we first present and elaborate the fiducial
simulation (see \S\ref{sec:fiducial-model}), then compare
the the results of the variant models shown in
\ref{table:various-model} (see \S\ref{sec:various-model}).

\subsection{Fiducial Model}
\label{sec:results-fiducial}

\begin{figure*}
  \centering
  \includegraphics[width=7.1in, keepaspectratio]
  {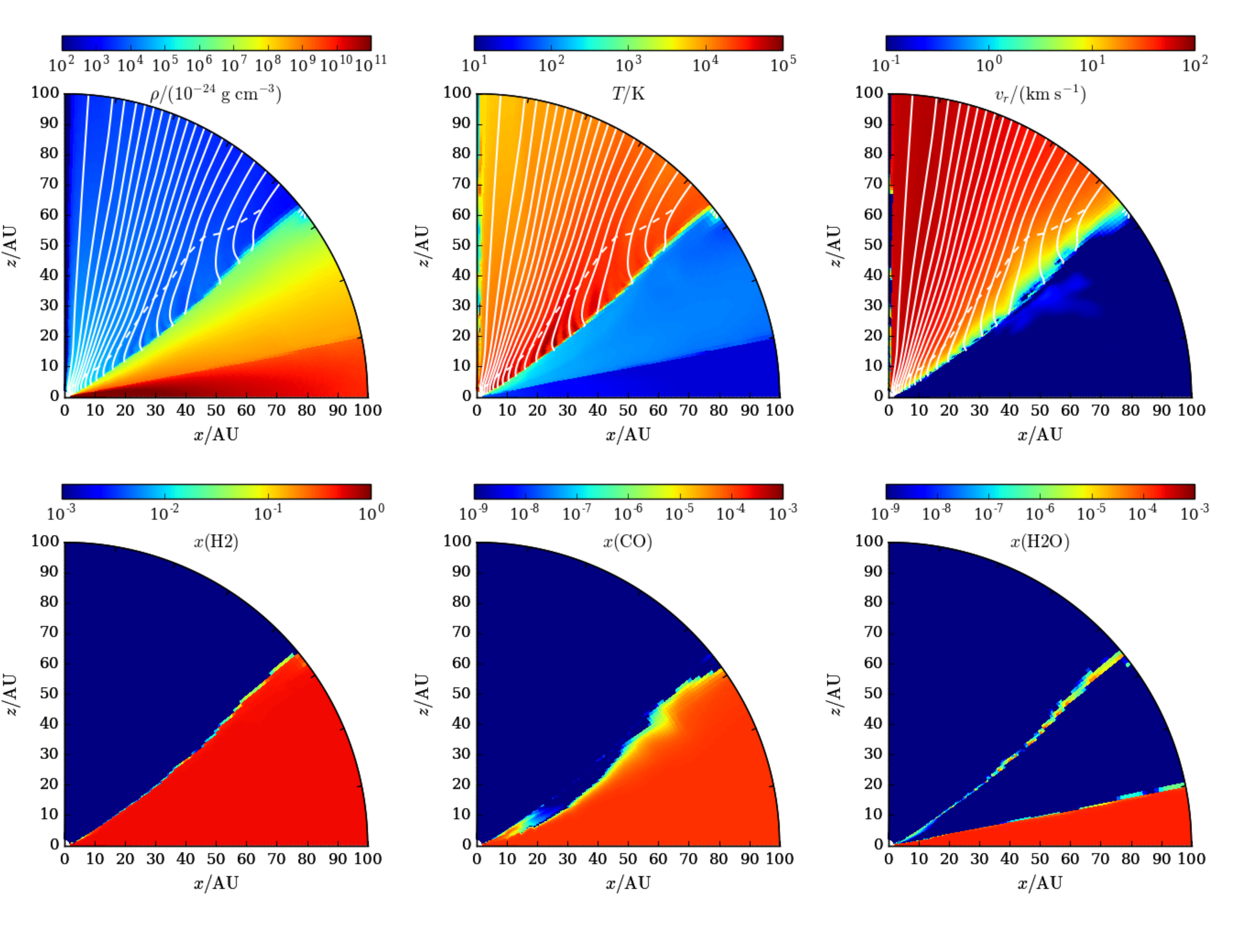}
  \caption{Meridional plots for the fiducial model
    (\S\ref{sec:fiducial-model}), averaged through the
    final $100~\yr$ of the simulation. {\bf Top row}: basic
    hydrodynamic profiles; left panel: mass density in units
    of $10^{-24}~\g~\cm^{-3}$; middle panel: temperature in
    Kelvin; right panel: radial velocity in $\kms$. Panels
    in the top row are overlapped by streamlines (white
    solid lines), separated by $10^{-10}~M_\odot~\yr^{-1}$
    wind mass loss rate (see \S\ref{sec:results-fiducial}
    for details), and the locations of sonic points (white
    dashed line). {\bf Bottom row}: relative abundance of
    different species [in units of
    $n(\chem{X})/n_{\chem{H}}$]; left panel: \chem{H_2};
    middle panel: \chem{CO}; right panel: \chem{H_2O}. }
  \label{fig:slice_fiducial}
\end{figure*}

Fig.~\ref{fig:slice_fiducial} displays meridional plots of
the structure of our fiducial model averaged over the final
$100~\yr$ of the simulation. The white curves shown in the
top row of panels are streamlines, the integral curves of
the vector field $\rho\b{v}_p$ ($\b{v}_p$ is the poloidal
velocity), spaced by constant mass-loss rate
$10^{-10}~M_\odot~\yr^{-1}$: that is to say, this is the
mass flux between neighboring streamlines when integrated
over azimuth and multiplied by two to include the reflection
of the computational region below the equatorial plane.
Streamlines that meet the outer boundary with a negative
value of the Bernoulli parameter
\begin{equation}\label{eq:Bern}
  \mathcal{B} \equiv \dfrac{v^2}{2} +
  \dfrac{\gamma p}{(\gamma-1)\rho} + \Phi\ ,
\end{equation}
are not plotted, and the outflow along such streamlines is
omitted from the computation of the total mass-loss rate.
Here $v$ is the magnitude of fluid velocity vector, $p$ the
gas pressure, $\gamma\approx 5/3$ the adiabatic index, and
$\Phi$ the gravitational potential. With this mask we get
rid of (very slow) radial flows near the mid-plane: since
the density there is six orders of magnitude higher than the
wind, a tiny radial velocity fluctuation could otherwise
give a spurious contribution to the mass-loss rate. As
displayed in Fig.~\ref{fig:slice_fiducial}, the streamlines
terminate on the disk at the surface where $\mathcal{B}$
becomes negative.  We consider this surface to be the base
of the wind.  (As discussed in \S\ref{sec:GH09_comparison},
this definition of the wind base differs from that of GH09.)

\begin{figure}
  \centering
  \includegraphics[width=3.3in, keepaspectratio]
  {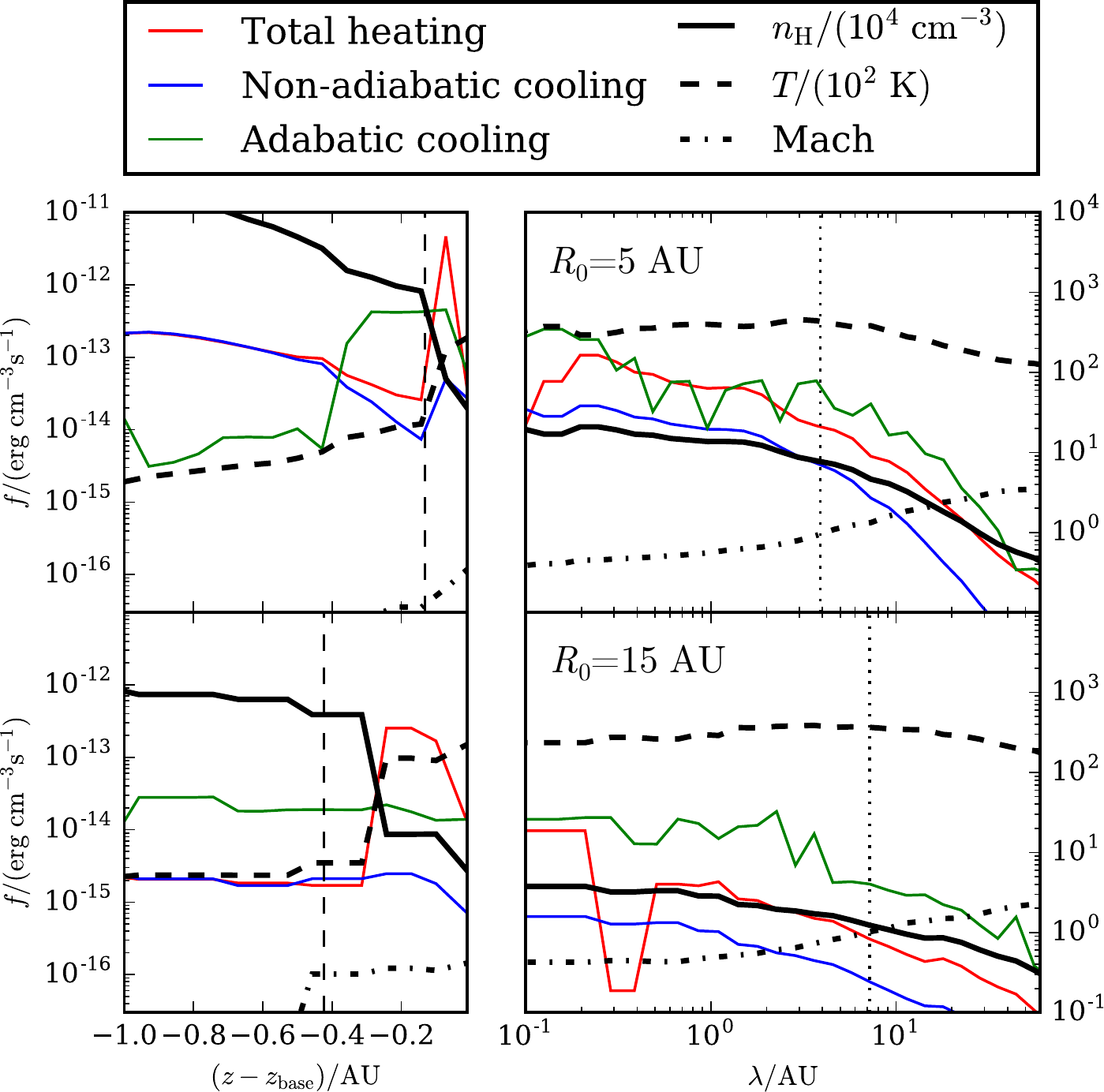}
  \caption{Cooling/heating rates (left ordinate) and fluid
    properties (right ordinate) along
    streamlines. $n_{\chem{H}}$ is the number density of
    {\it hydrogen nuclei}. In right column, horizontal axis
    is arc length ($\lambda$) measured from the wind base
    [where $\mathcal{B}=0$, eq.~\eqref{eq:Bern}]. Vertical
    dotted line marks sonic point. Left column present the
    profile {\it vertically below} wind base (i.e.
    $z - z_\mathrm{base}$).  Vertical dashed line indicates
    wind base as defined by GH09.  Upper row: streamline
    rooted at $R_0=5~\au$.  Lower row: $R_0=15~\au$.  }
  \label{fig:streamline}
\end{figure}

Fig.~\ref{fig:streamline} shows several flow variables along
two representative streamlines originating from cylindrical
radii $R\equiv r\sin\theta=5~\au$ and $R=15~\au$.

The density and temperature profiles shown by
Fig.~\ref{fig:slice_fiducial} can be divided into three
relatively distinct regions:
\begin{itemize}
\item Midplane layer: $0 < (z/R) \lesssim 0.3$
  ($R=r\sin\theta$ being cylindrical radius), $T<10^2\K$.
  The structure here is basically unchanged from the initial
  conditions.
\item Intermediate layer: $0.3\lesssim (z/R) \lesssim 0.6$,
  $10^{-19}~\g~\cm^{-3} \lesssim \rho \lesssim
  10^{-16}~\g~\cm^{-3}$, $10^2\lesssim T\lesssim
  10^3~\K$. The total mass in this layer is
  $\sim 10^{-6}~M_\odot$. EUV photons scarcely penetrate
  this region, whose properties are controlled by FUV and
  X-ray processes: photodissociation and photoelectric
  heating, as well as radiative cooling by collisionally
  excited molecular and/or atomic transitions. Most
  \chem{H_2} molecules and a lot of \chem{CO} molecules
  survive in this region because of significant self- and
  cross-shielding of Lyman-Werner photons. Soft FUV photons
  that do not interact much with molecular hydrogen are
  relatively unshielded and pervade the intermediate layer,
  photodissociating \chem{OH} and \chem{H_2O}), penetrating
  to the bottom of the layer, or escaping through the outer
  radial boundary.
\item Wind layer: $(z/R)\gtrsim 0.6$,
  $\rho \lesssim 10^{-19}~\g~\cm^{-3}$, $T \gtrsim
  10^4~\K$. This region is filled with mostly ionized gas,
  flowing outwards at radial velocity $v_r\sim
  30~\kms$. Photoionization heating and adiabatic expansion
  dominate the thermodynamics of this region.
\end{itemize}

\begin{figure}
  \centering
  \includegraphics[width=2.8in, keepaspectratio]
  {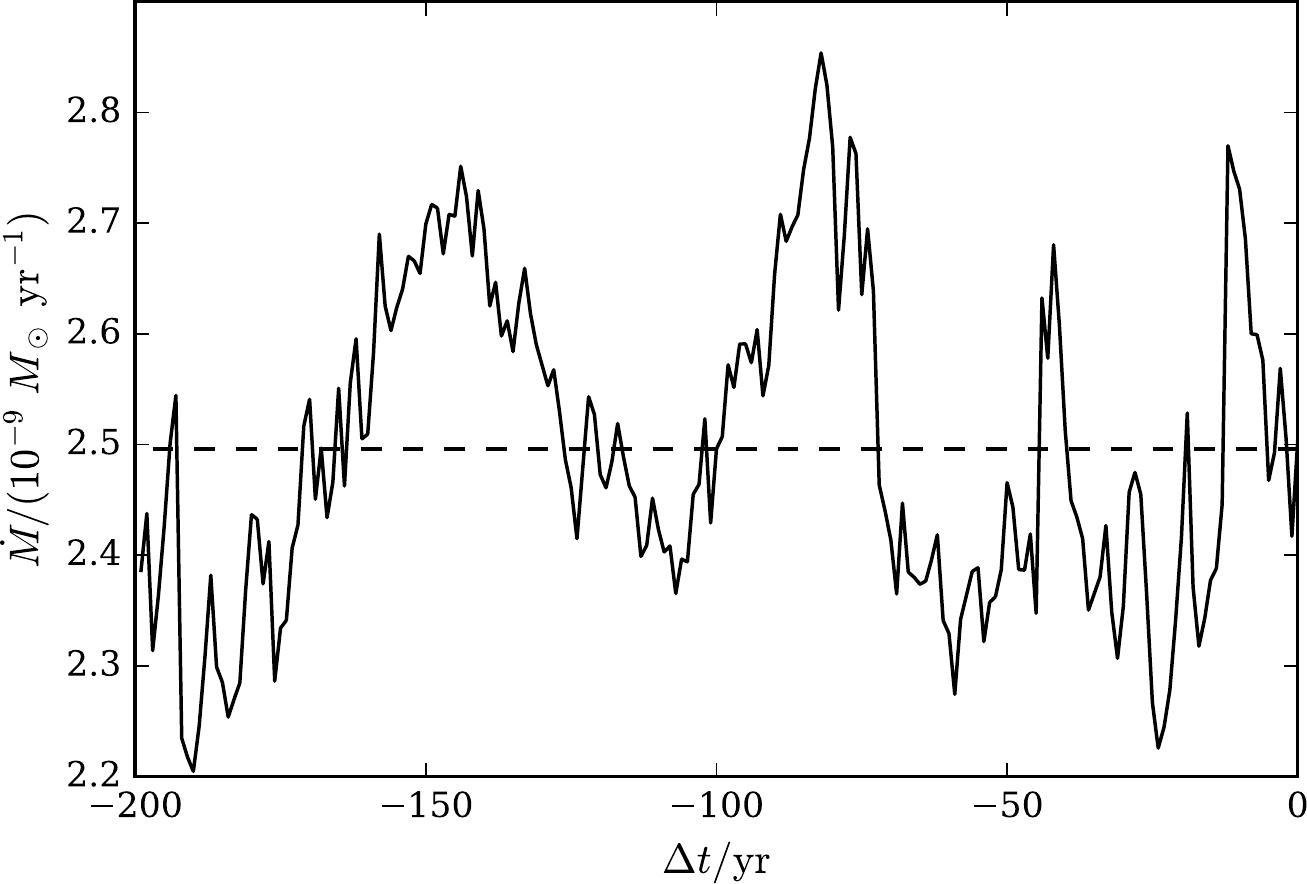}
  \caption{Variation of mass-loss rate measured at the
    $r=100~\au$ outer boundary of the fiducial model
    (Model~0). The dashed horizontal line shows the average
    of mass-loss rate over the last $200~\yr$.}
  \label{fig:mdot_var}
\end{figure}

If we integrate the $\mathcal{B}$-masked radial mass flux at
the the $r=100~\au$ boundary (and its reflection at $z<0$)
and average over the last $100~\yr$ of our fiducial run, we
obtain a total mass-loss rate
$\dot{M}_\wind \simeq 3.4\times 10^{-9}~M_\odot~\yr^{-1}$,
corresponding to a disk dispersal timescale $\sim
10^7~\yr$. The mass-loss rate is lower than that of GH09
(see \S\ref{sec:GH09_comparison} for further discussion).
However, our mass-loss rate undergoes significant
fluctuations, and is uncertain to at least $\sim 10$ per cent.
Fig.~\ref{fig:mdot_var} plots the mass-loss rate for the
last $200~\yr$ of the (lower resolution) fiducial run
(Model~0). They correlate with what appears to be a thermal
instability of the outer disk, whereby it swells vertically,
intercepts more radiation, and then swells further but also
migrates at a few $\kms$ through the outer boundary,
temporarily increasing $\dot{M}$.  This behavior is smoothed
over by the time averages used to make
Fig.~\ref{fig:slice_fiducial}.  These swellings, being
slower and denser than the general wind, partly shield
themselves against photodissociation of some molecules,
especially \chem{H_2} and \chem{CO}, so that those molecules
survive farther into the outflow than they would otherwise.

Even outside these swellings, there are also molecules
surviving in regions with rather high temperature
($\sim 10^3~\K$, or even up to $\sim 10^4~\K$). \chem{H_2O}
and \chem{OH} molecules exist at the surface of the
intermediate layer, detached from the midplane (last panel
of Fig.~\ref{fig:slice_fiducial}).  The reformation rates of
\chem{H_2O} and \chem{OH} are comparable to
photodissociation at that surface.  At the cooler
temperatures below it, inside the intermediate layer,
reformation is less efficient but photodissociating FUV is
still present.  The wind region, on the other hand, does not
have sufficient \chem{H_2} (reactions that are most
efficient in forming \chem{H_2O} and \chem{OH} need
\chem{H_2} as reactants, while the reactions that convert
atomic H to OH and \chem{H_2O} are very slow).

\begin{figure*}
  \centering
  \includegraphics[width=7.1in, keepaspectratio]
  {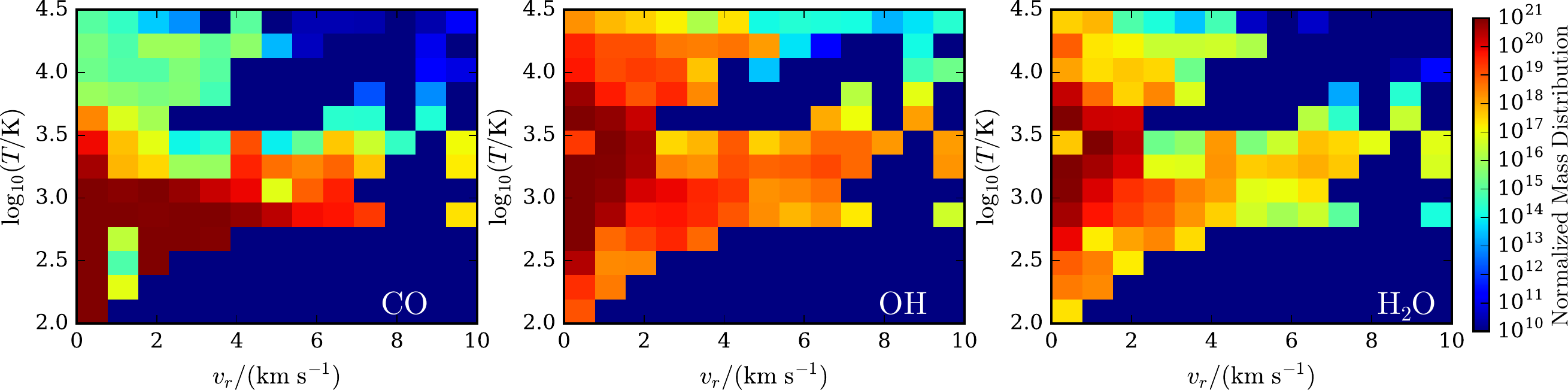}
  \caption{Distribution functions of key molecular species
    based on the fiducial model (left panel: CO; middle
    panel: OH; right panel: \chem{H_2O}) in the
    two-dimensional space of $\{\log_{10}T\}\times\{v_r\}$
    (common logarithm of temperature and radial
    velocity). The mass distribution function is normalized
    as $\d^2m / [\d\log_{10}(T/\K)\d(v_r/\kms)]$, i.e. mass
    of the species per dex-temperature per $(v_r/\kms)$. }
  \label{fig:phase_species}
\end{figure*}

In Fig.~\ref{fig:phase_species}, we plot the distribution of
\chem{CO}, \chem{OH} and \chem{H_2O} in the wind region and
intermediate layer, in the plane of by $\log_{10}T$ and
$v_r$. (The temperature $T$ here represents the kinetic
temperature of the local (mostly $\chem{H}$ \& $\chem{He}$)
gas, not the vibrational or even rotational excitation
temperature of the molecules.) For those molecules, a tail
on the high temperature ($T\sim 10^3\dash 10^4~\K$) and
intermediate radial velocity ($v_r\sim 5\dash 10~\kms$) end
of the 2-D distribution indicate their survival at the
bottom in the wind region. Such hot molecular gas would be
less prominent had we assumed local thermochemical
equilibrium. For the luminosity in the LW and EUV bands of
our fiducial model, it can be estimated that the timescale
of CO photodissociation is $\sim 0.1~\yr$ at $r\sim
10~\au$. Given the speed of photoevaporative outflow, this
timescale is sufficient for some CO to survive
$\sim 0.1-1~\au$ into the hot wind.  These timescales are
sensitive to radial distance (from the radiation sources),
to the way photodissociation is modeled (see
\S\ref{sec:chem-thermo}), and to the LW and EUV band
luminosity. Observational constraints on such molecules
could be an important check on these models, and might
diagnose the role of UV in driving PPD winds.

\subsection{Exploring the Parameter Space}
\label{sec:various-model}

To explore the effects of our input parameters, we have run
a number of additional simulations, most differing from the
fiducial run in one parameter.  These models and some synoptic
results are listed in
Table~\ref{table:various-model}.
We discuss some of these models here, and others in
\S\S\ref{sec:GH09_comparison}-\ref{sec:compare-oeca10}
in relation to the works by GH09 and OECA10.

In the fiducial model the luminosity in the Lyman-Werner
band is tiny compared to that in soft ($h\nu < 11.3~\eV$)
FUV photons: around 0.35 per cent, using the $9000~\K$ black
body SED. However, as observed by
e.g. \citet{2000ApJ...544..927G}, the SED for FUV radiation
is rather variable from object to object and often more
luminous in the LW band than the black-body model adopted by
GH08 and GH09. Hence we include a series of models, with 0
and 100 times the fiducial luminosity in the Lyman-Werner
band, to cover this uncertainty. We also test 0 and 10 times
EUV or X-ray luminosity to diagnose the impact of those
photons that can ionize atomic hydrogen.

For very small grains such as our PAHs, the grain absorption
cross section for FUV and EUV photons depends on total grain
mass rather than grain area.  We have a much smaller grain
mass than GH09.  Model~9 in Table~\ref{table:various-model}
has double the dust radius ($r_\dust = 10~\ang$) and
therefore eight times the dust mass at the same relative
number density ($n_{\chem{Gr}}/n_{\chem{H}}=10^{-7}$).

To test our truncation errors, we repeat the fiducial run at
resolution $128\times 64$, i.e. coarser by $\times 2$ in
both latitude and radius.  This convergence test is run for
much longer time period ($\sim 2000~\yr$) to better
characterize fluctuations around the mean state.

\begin{figure*}
  \centering
  \includegraphics[width=7.1in, keepaspectratio]
  {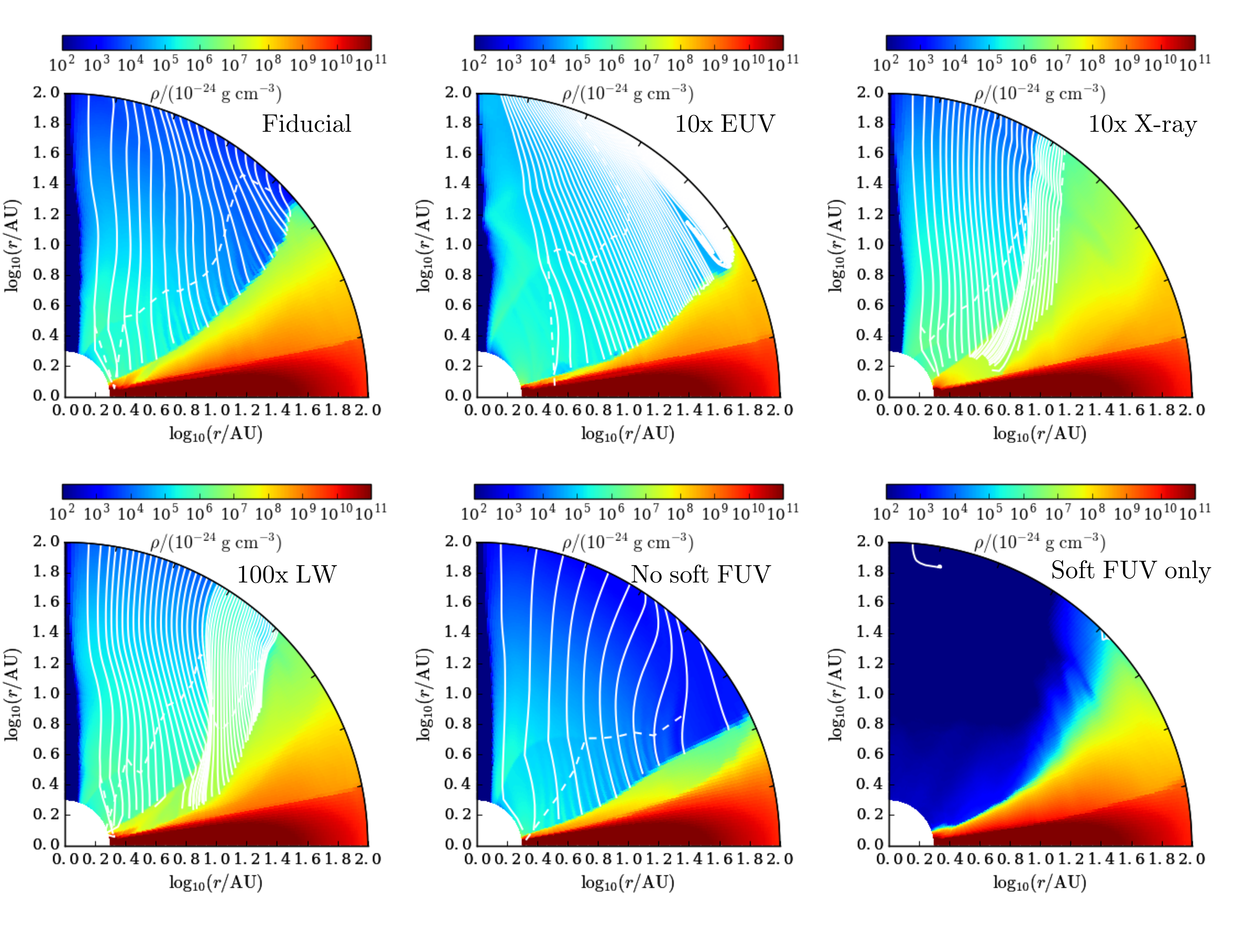}
  \caption{Selected plots of streamlines (based on $100~\yr$
    average), overlaid on density colormaps, that are
    relevant to the discussions in \S\ref{sec:results}. The
    black curves, showing the streamlines, are separated by
    $10^{-10}~M_\odot~\yr^{-1}$ each. The white curves
    denote the location of sonic point of all streamlines
    plotted. Note that, for clearer presentation, the radial
    coordinates in all panels are $\log_{10}(r/\au)$.}
  \label{fig:flux-examples}
\end{figure*}

Fig.~\ref{fig:flux-examples} illustrates the hydrodynamic
structure of a few representative models. These plots are
based on time averages over $100~\yr$, so that the flow
field is in approximate steady state.  In the runs with
$100\times$ LW photons (Model 2) or $10\times$ X-ray photons
(Model 8), a thick neutral atomic layer exists at the top of
the intermediate layer. In this layer, the temperature and
sound speed reach a local maximum with respect to height or
latitude, and significant outflows may occur.  This causes
the jagged shape of the sonic curves in the third and fourth
panels.

\begin{deluxetable*}{ccccccc}
\tablecolumns{10}
\tabletypesize{\scriptsize}
\tablewidth{0pt}
\tablecaption{Models exploring parameter space}
\tablehead{
  \colhead{No.} &
  \colhead{Description} &
  \colhead{$\dot{M}_\wind$} &
  \colhead{$\dot{M}_{\mathrm{GH}}$} &
  \colhead{Total heating} &
  \colhead{Efficiency} &
  \colhead{$\langle v_r\rangle$}
  \\
  \colhead{} &
  \colhead{} & 
  \colhead{$(10^{-9}M_\odot~\yr^{-1})$} &
  \colhead{$(10^{-9}M_\odot~\yr^{-1})$} &
  \colhead{$(10^{30}\erg~\s^{-1})$} &
  \colhead{} &
  \colhead{$(\kms)$}
  \\
  \colhead{(1)} & \colhead{(2)} & \colhead{(3)} & 
  \colhead{(4)} & \colhead{(5)} & \colhead{(6)} & 
  \colhead{(7)} 
}
\startdata
0 & Fiducial  & 2.5 $\pm$ 0.2 & 11.6 & 4.4 & 0.67 & 39 \\
1 & No LW photons  & 2.5 $\pm$ 0.3 & 9.3 & 4.1 & 0.67 & 38 \\
2 & 100$\times$ LW photons  & 17.6 $\pm$ 2.1 & 61.3 & 9.1 &
0.60 & 18 \\ 
3 & No "soft" FUV  & 1.1 $\pm$ 0.1 & 2.7 & 2.3 & 0.53 & 58 \\
4 & "Soft" FUV only  & 0.0 & 0.2 & 1.0 & - & - \\
5 & No EUV  & 0.0 & 3.7 & 1.6 & - & - \\
6 & 10$\times$ EUV photons  & 9.4 $\pm$ 0.7 & 107.8 & 26.7 &
0.74 & 33 \\ 
7 & No X-ray  & 2.1 $\pm$ 0.2 & 6.9 & 2.6 & 0.80 & 38 \\
8 & 10$\times$ X-ray photons  & 9.1 $\pm$ 0.4 & 55.6 & 14.1
& 0.42 & 24 \\ 
9 & $r_\mathrm{dust}=10~\mathrm{\AA}$  & 2.8 $\pm$ 0.6 &
10.5 & 3.5 & 0.68 & 30 \\ 
10 & $\mathrm{OECA10\ analog}^\dagger$ & 11.2 $\pm$ 4.2 & 105.0 &
0.8 & 0.54 & 5 \\   
11 & Convergence test  & 2.7 $\pm$ 0.6 & 16.1 & 3.0 & 0.58 &
32 \\  
\enddata
\tablecomments{
  (1) Model identifier.
  (2) Parameter by which model differs from fiducial.
  (3) Wind mass-loss rate.  The error quoted error is
  $\Delta \dot{M}_\wind = \langle[\dot{M}(t)]^2
  - \langle \dot{M}\rangle^2 \rangle^{1/2}$, 
  where the time averages are taken over the last $100~\yr$.
  (4) Estimated wind mass-loss rate using GH09 scheme.
  (5) Total radiative plus thermal-accomodation
  heating of the gas (note that the accomodation
  heating can be negative).
  (6) Thermal-to-mechanical conversion efficiency:
  (heating $-$ non-adiabatic cooling)$/$(heating).
  (7) Mean outflow velocity weighted by radial mass flux.\\
  $\dagger$: Bernoulli parameter mask not applied;
  significant outflow occurs in the intermediate layer with
  $\mathcal{B} < 0$.
}
\label{table:various-model}
\end{deluxetable*}

Figs.~\ref{fig:heating} and
\ref{fig:cooling} show the vertical distributions of
heating and cooling mechanisms at $R=15~\au$, a typical
location where the outflow streamlines originate. The
three layer structure (\S\ref{sec:results-fiducial}) is
obvious in most of the models. Details of those structures
vary with model parameters, with implications for the
mechanisms responsible.

The panels of Fig.~\ref{fig:heating} convey some general
impressions about the heating mechanisms.  The vertical
heating profile usually has two peaks: one at the bottom of
the intermediate layer, the other at the top of it.
Photoionization heating by the harder (EUV and
X-ray) photons dominates, unless these photons are absent or
are overwhelmed by photons in other bands (e.g. Model 2,
$100\times$ LW photons; see discussions below). On the
cooling side (Fig.~\ref{fig:cooling}), the \chem{OH} and/or
\chem{H_2O} ro-vibrational transitions and S I $25~\micron$
transition dominate at the bottom of intermediate layer, the
Si II $35~\micron$ and O I $63~\micron$ transitions in the
middle of that layer, and \chem{H_2} ro-vibrational cooling
near the top. In the ``wind'' region, cooling and heating
are dominated by recombination and the photoelectric
effect.   Using the integrated cooling rate,
we have estimated some of the important
line luminosities (Table~\ref{table:line-luminosity}).

\begin{deluxetable*}{cccccccc}[h]
  \tablecolumns{8} 
  \tabletypesize{\scriptsize}
  \tablewidth{0pt}
  \tablecaption{Approximate line luminosity}   
  \tablehead{
    \colhead{Model} &
    \colhead{O I $63\ \micron$} &
    \colhead{O I $6300~\ang$} &
    \colhead{S I $25\ \micron$} &
    \colhead{Si II $35\ \micron$} &
    \colhead{\chem{H_2} ro-vib} &
    \colhead{\chem{OH}/\chem{H_2O} ro-vib} &
    \colhead{\chem{CO} ro-vib}
  }
  \startdata
  0  & $-4.31$ & $-7.36$ & $-4.00$ & $-5.65$ & $-4.67$ &
  $-4.29$ & $-3.89$ \\ 
  1  & $-4.34$ & $-8.17$ & $-3.96$ & $-5.73$ & $-4.60$ &
  $-4.36$ & $-3.90$ \\ 
  2  & $-3.75$ & $-4.69$ & $-4.17$ & $-4.73$ & $-4.11$ &
  $-3.57$ & $-3.91$ \\ 
  3  & $-5.05$ & $-7.72$ & $-5.05$ & $-6.48$ & $-5.26$ &
  $-3.61$ & $-5.68$ \\ 
  4 & $-5.05$ & - & $-4.68$ & - & $-5.61$ & $-4.30$ &
  $-4.26$ \\ 
  5  & $-4.22$ & $-7.31$ & $-3.97$ & $-5.58$ & $-5.03$ &
  $-4.63$ & $-3.88$ \\ 
  6  & $-4.43$ & $-7.13$ & $-3.88$ & $-5.36$ & $-3.65$ &
  $-3.56$ & $-3.78$ \\ 
  7  & $-4.74$ & $-8.02$ & $-4.61$ & $-6.30$ & $-4.72$ &
  $-4.26$ & $-4.23$ \\ 
  8  & $-3.57$ & $-5.99$ & $-3.11$ & $-4.69$ & $-4.15$ &
  $-3.38$ & $-3.37$ \\ 
  9  & $-4.47$ & $-4.94$ & $-4.39$ & $-5.94$ & $-4.39$ &
  $-4.07$ & $-4.59$ \\ 
  10  & $-4.38$ & $-6.88$ & - & $-4.55$ & - & - & - \\
  11  & $-4.49$ & $-5.24$ & $-4.32$ & $-5.98$ & $-4.79$ &
  $-4.94$ & $-4.83$ \\ 
  \tablecomments{All luminosities are presented in
    $\log_{10}(L/L_\odot)$.}
\label{table:line-luminosity}
\end{deluxetable*}

\section{Discussion}
\label{sec:discussions}

\subsection{Roles of different bands of radiation}

The heating and thickness of the intermediate layer is
mainly attributable to photons that penetrate a radial
column
$\gtrsim 10^{21}~\cm^{-2}\sim 10^8~\cm^{-3}\times 10~\au$,
viz. soft FUV and X-rays. These two components penetrate to
$n\sim 10^8~\cm^{-3}$, which approximately defines the top
of the midplane layer.  Scattered $\lya$ photons would
penetrate to vertical columns $\sim 10^{22}\cm^{-2}$ if they
were included (see \citealt{2011ApJ...739...78B}).

Heating processes inside the intermediate layer do not
directly contribute to the outflows.  This is suggested in
Fig.~\ref{fig:slice_fiducial} by the terminations of the
streamlines, which mark the surface where $\mathcal{B}=0$,
at the top of the intermediate layer; and also by the
mass-loss rates in Table~\ref{table:various-model},
especially for models 5 (soft FUV only), 6 (no EUV photons),
and and 3 (no soft FUV).  The relatively high density in the
bulk of the intermediate layer ($n \gtrsim 10^6\cm^{-3}$)
causes cooling processes to offset much of the radiative
heating there. Nevertheless, the FUV photons, and in most
cases the X-ray photons also, contribute indirectly to the
mass-loss rate by thickening the intermediate layer, which
exposes its upper surface to more intense EUV heating.

The LW photons play a more direct role. Due to
self-shielding of \chem{H_2}, heating by LW photons extends
only slightly below the upper surface of the intermediate
layer. These photons contribute dramatically to the total
outflow. Admittedly, this is not convincingly shown by
Table~\ref{table:various-model} alone: compare Models~0
(fiducial), 1 (no LW photons) and 2 ($10\times$ LW photons).
In these three cases, the luminosities in the LW band are
smaller than the EUV component.  But the corresponding
panels of~Fig.~\ref{fig:heating} reveal a clear trend of
increasing heating by \chem{H_2} pumping (see Appendix
\ref{sec:fuv-h2}).  Model 2 ($100\times$ LW photons) clearly
demonstrates that LW heating by interaction with \chem{H_2}
can dominate the outflow when the LW is sufficient.  It is
worth noting that, for Model 2, the outflow velocity is
approximately $\sim 1/3$ that of Models 0, 1, and 2: LW
photons deposit less energy per reaction than EUV and X-ray
(see also \S\ref{sec:chem-thermo} and Appendix
\ref{sec:fuv-h2}), thereby heating the gas to a lower
temperature and hence accelerating the outflow to a lower
velocity.

EUV photons are absorbed by a rather small column density
($\sim 10^{18}~\cm^{-2}$) of neutral or molecular hydrogen.
However, if they make their way to the surface of
intermediate layer, EUV photons tend to dominate the
photoevaporative outflow. This point is illustrated by
comparison of Model 0 (fiducial) with 6 (no EUV) and 7
($10\times$ EUV): Model 5 has its outflow almost totally
shut down, while Model 6 has a dramatically increased mass
loss rate and radial flow velocity.  Because the EUV heating
is concentrated in relatively little mass, it produces a
high sound speed and therefore a relatively fast wind.

Since the X-rays have considerably larger penetration than
LW and EUV photons, in fact comparable to that of the softer
FUV photons, they deposit most of their heat in the
intermediate layer, where cooling mechanisms remove energy
efficiently from the gas. The exception among the cases in
Table~\ref{table:various-model} is Model 8 ($10\times$
X-ray), where X-rays drive a substantial wind by themselves.
Note that the wind velocity is about half that of the
fiducial model because the heating is distributed over a
larger mass.  X-ray driving is further discussed below in
connection with the work of \citet{2010MNRAS.401.1415O}.

\begin{figure*}
  \centering
  \includegraphics[width=7.1in, keepaspectratio]
  {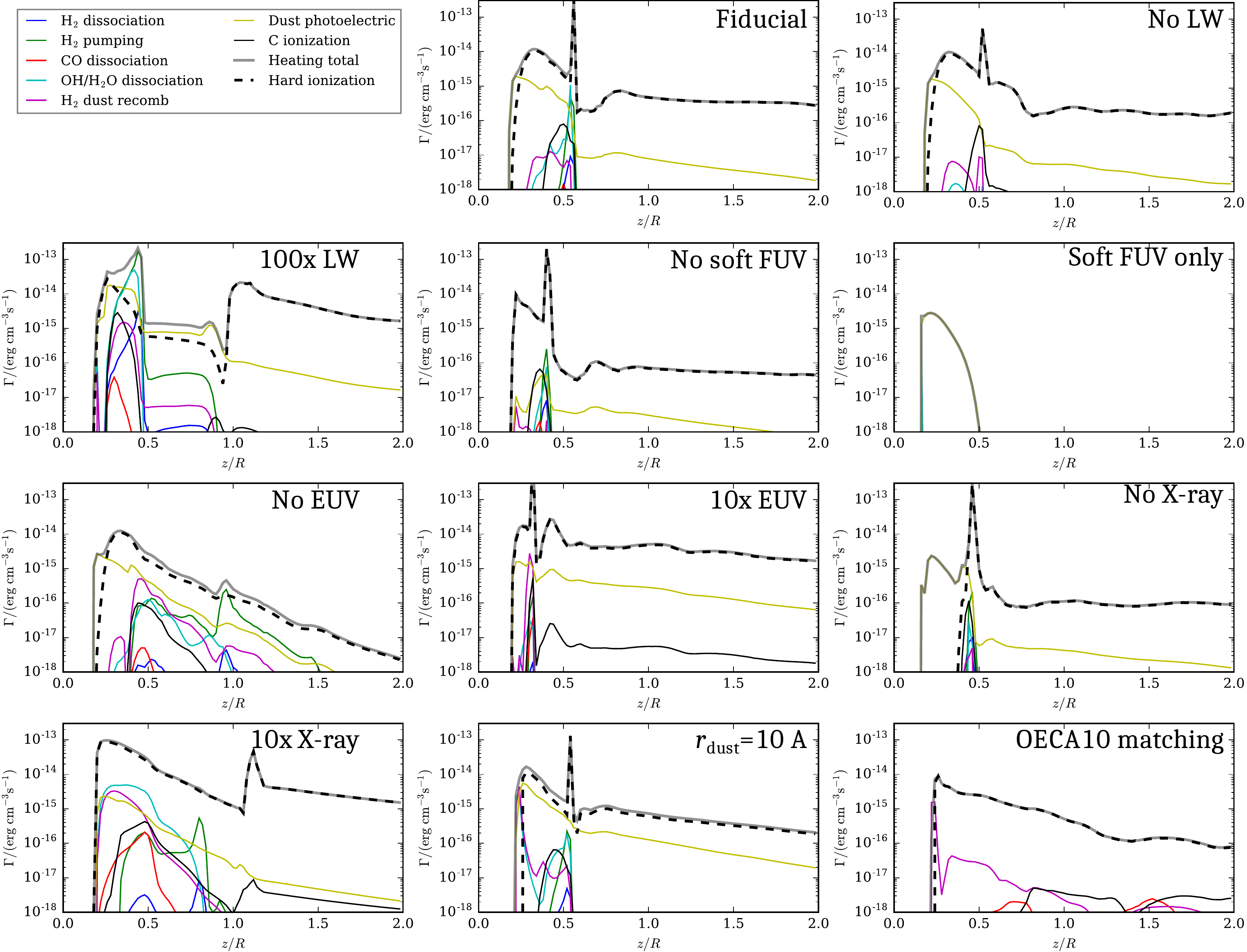}
  \caption{Profiles of heating mechanisms, in units of
    $\erg~\cm^{-3}~\s^{-1}$, at $R=15~\au$, for all models
    involved in this paper.  Heating mechanisms are
    distinguished by line shape and color as marked
    in the legend at the upper left.}
  \label{fig:heating}
\end{figure*}

\begin{figure*}
  \centering
  \includegraphics[width=7.1in, keepaspectratio]
  {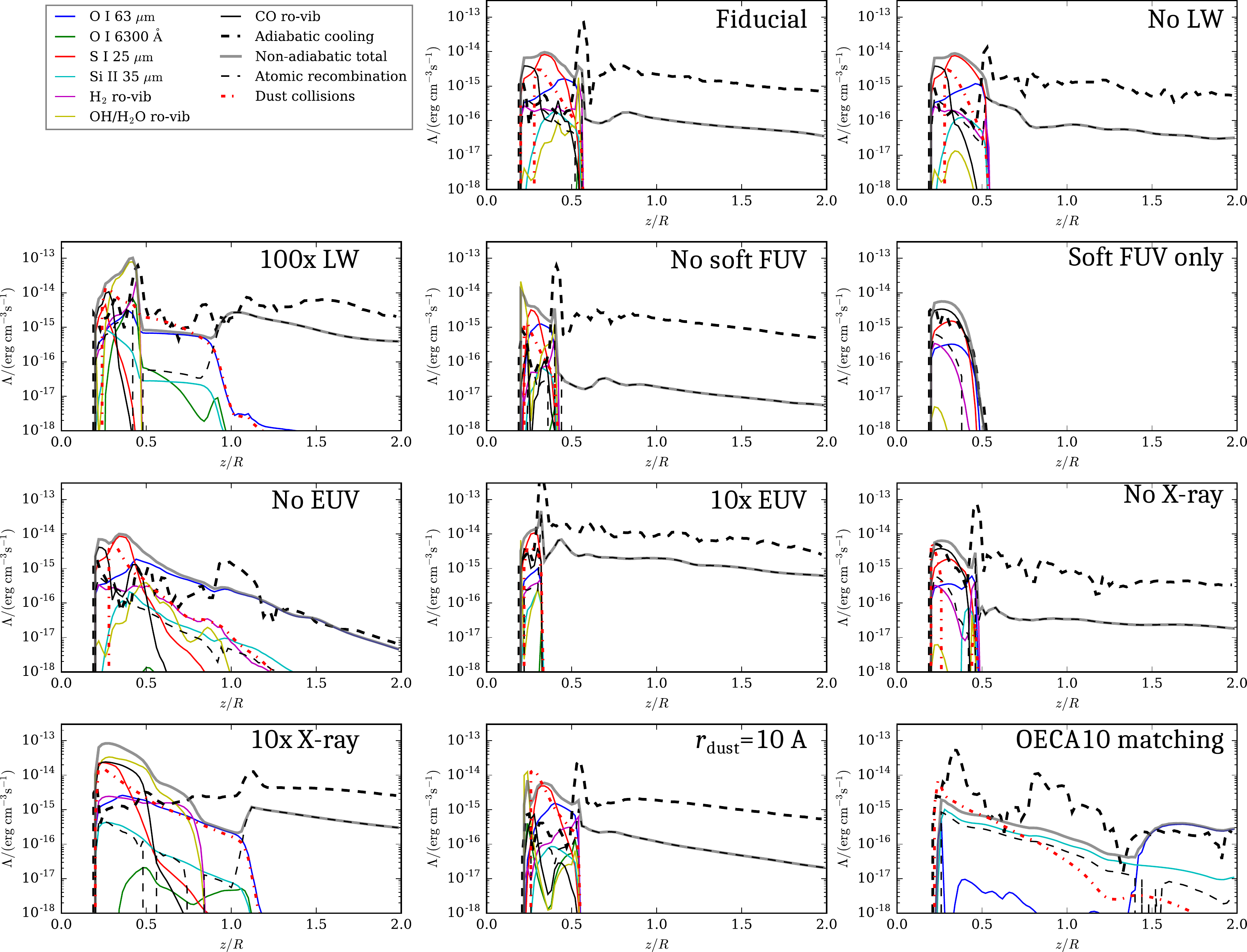}
  \caption{Same as Fig.~\ref{fig:heating}, but shows
    profiles of different cooling mechanisms at $R=15~\au$.}
  \label{fig:cooling}
\end{figure*}

\subsection{Comparison with GH09}
\label{sec:GH09_comparison}

\begin{figure}
  \centering
  \includegraphics[width=2.8in, keepaspectratio]
  {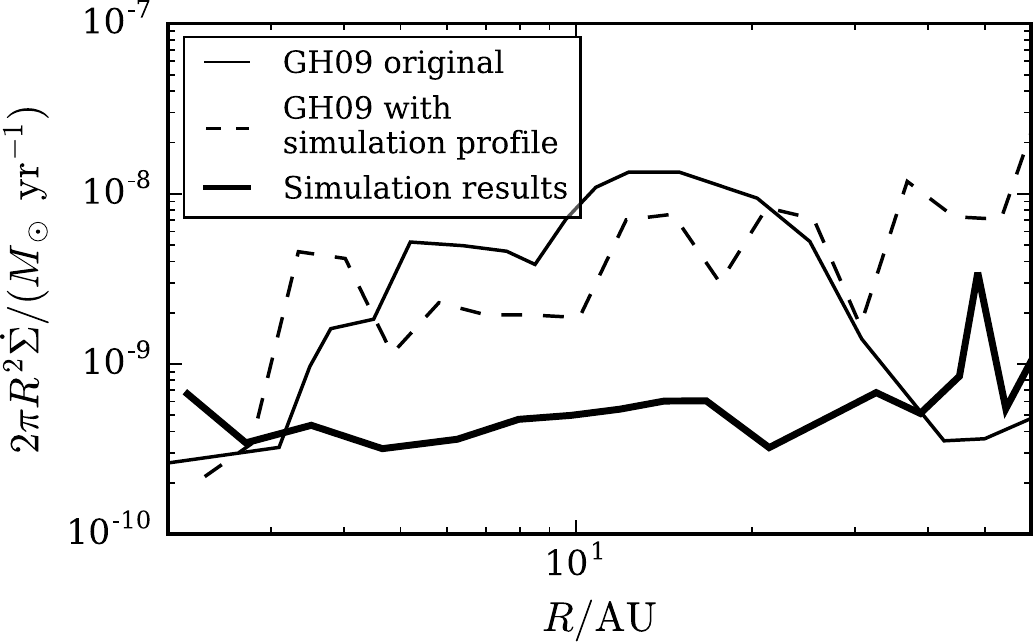}
  \caption{Radial profiles of the differential wind mass
    loss rate, $(2\pi R^2\dot{\Sigma})$, for the
    time-averaged fiducial simulation.  Dashed curve is
    calculated by applying GH09's prescription to the
    density and temperature structure of our simulation,
    whereas solid curve is our actual mass-loss rate.}
  \label{fig:radial_mdot}
\end{figure}

Although we have modeled our fiducial case on that of GH09,
our mass-loss rate is a few times smaller than theirs.
GH09 did not simulate the hydrodynamic flow in two
dimensions as we have done, but instead estimated $\dot M$
by analytically matching their heated disks onto a spherical
Parker wind.  The fourth column of
Table~\ref{table:various-model} shows the rates that would
be estimated by applying their prescription to our heated
disks.  For the fiducial case, this estimate matches GH09's
results fairly well, as determined by integrating their
radius-dependent mass-loss rate from 1 through 100
$\au$. The latter agreement suggests that our simplified
thermochemical network is sufficient to predict the
temperature, density, and flow, if not all of the trace
species and line emission that one would like to compare
with observations.  The differences between the third and
fourth columns point to the importance of modeling the
hydrodynamics properly, however.
Fig.~\ref{fig:radial_mdot} shows the radial profiles of
the mass-loss computed in these two different ways, and also
for GH09's original model disk.  Note that the latter
extended beyond $100~\au$.

GH09's analytic prescription computes the mass-loss rate as
if the disk atmosphere, which in their models is
hydrostatic, belonged to a spherical Parker wind with
constant sound speed, i.e. an adiabatic wind with an
isothermal equation of state.  For each cylindrical radius
$R$, the matching point between the hydrostatic atmosphere
and the isothermal wind are chosen at such an altitude as to
maximize the imputed mass-loss rate.  This is how GH09
determine the wind base, $z_{\rm b}(R)$.  The sound speed of
the wind is then the sound speed of the hydrostatic
atmosphere at $z=z_b$.  Note that $\dot M$---or rather
$2\pi R^2\dot\Sigma(R)$---can be expected to have such a
maximum with respect to the matching altitude $z_b$ because
of the strong vertical gradients of temperature and density
within the atmosphere.

To understand better why our mass-loss rates differ from
GH09's prescription, consider the streamline starting from
$R_0=15~\au$ in Fig.~\ref{fig:slice_fiducial} (see also the
lower panel of Fig.~\ref{fig:streamline}) .  Using the
density and temperature profiles obtained by our fiducial
simulation, the density and temperature at the wind base
(defined in the manner of GH09) read
$n_{\chem{H}} \simeq 3.7\times 10^6~\cm^{-3}$ and
$T\simeq 0.7\times 10^3~\K$.  These values are in good
agreement with figure 1 in GH09.  The corresponding
isothermal sound speed is $\cs\simeq 2.1~\kms$: the hydrogen
is atomic and neutral, with mean mass per particle
$\bar m \simeq 1.3m_{\chem{H}}$.  The implied sonic radius
is then $r_s \simeq 78~\au$, and the predicted density
there is $n_{\chem{H},s}\simeq 10^5~\cm^{-3}$.  The
corresponding mass-loss rate for a spherical isothermal wind
with these sonic-point parameters would be
$\dot M \sim 2\pi r_s^2 \bar m n_{\chem{H},s} \cs \simeq
7\times 10^{-9}~M_\odot~\yr^{-1}$.  This is identified with
the local mass-loss rate per logarithmic radius at the
surface of the disk, viz. $2\pi R^2\dot{\Sigma}$ .
  
However, in the simulation, the actual sonic point on this
streamline lies at $r_s \simeq 23~\au$.  The actual density
there is $n_{\chem{H},s}\simeq 1.2\times 10^4~\cm^{-3}$, and
the temperature is nearly $3.6\times 10^4~\K$, so that the
gas is largely ionized ($\bar{m} \simeq
0.64~m_{\chem{H}}$). The isothermal sound speed based on
this temperature and molecular weight is
$\cs\simeq 20~\kms$.  The corresponding
$\dot M\approx 0.8\times 10^{-9}~M_\odot~\yr^{-1}$ (again,
this is indeed $2\pi R^2\dot{\Sigma}$).  This is about one
order of magnitude less than GH09's prescription.

In short, the actual mass-loss rate is smaller than GH09's
prescription by approximately the reciprocal of the ratio of
sound speeds (and hence flow velocities): $2.1~\kms$
vs. $20~\kms$.  The wind thrusts---momentum flux
$2\rho_s \cs^2$ times $4\pi r_s^2$---are nearly equal.  The
thrust is necessarily limited by the pressure of the gas at
the wind base, since the disk must support the force exerted
on it by the escaping wind.  Although we and GH09 define the
wind bases separately, the pressures are similar because the
bases are separated vertically by only $\sim0.5 \au$ out of
$\sim 9\au$. An analogy can be made here with rocketry: at
fixed thrust, the mass-flow rate in the rocket exhaust is
inversely proportional to the exhaust velocity, or
equivalently, to the specific impulse.  By assuming a
constant sound speed from their wind base upward, GH09's
prescription fails to account for the increase in specific
impulse due to the sharp rise in temperature above the wind
base, even though this rise is also seen in their own
hydrostatic disk atmospheres (see Fig.~1 of GH08).

\subsection{Comparison with OECA10}
\label{sec:compare-oeca10}

GH09 found that X-rays made a significant but not dominant
contribution to the mass-loss rate. OECA10, on the other
hand, modeled photoevaporation driven solely by X-rays, with
a full axisymmetric hydrodynamic treatment similar to our
own, though with a simplified prescription for the gas
temperature.  OECA10 assumed an X-ray luminosity close to
that of our fiducial model, without FUV or EUV, but found a
higher a wind mass-loss rate
$\sim 1.4\times 10^{-8}~M_\odot~\yr^{-1}$.  Although our
fiducial model has comparable X-ray and EUV luminosities,
($L_X=10^{30.4}~\erg~\s^{-1}$,
$L_{EUV}=10^{30.3}~\erg~s^{-1}$) EUV appears to dominate
the mass loss.  When we turn off the EUV, our mass-loss rate
becomes less than $10^{-10}~M_\odot~\yr^{-1}$ (Model~5 in
Table~\ref{table:various-model}).  We have investigated the
causes of these large differences.

\begin{figure}
  \centering
  \includegraphics[width=3.1in, keepaspectratio]
  {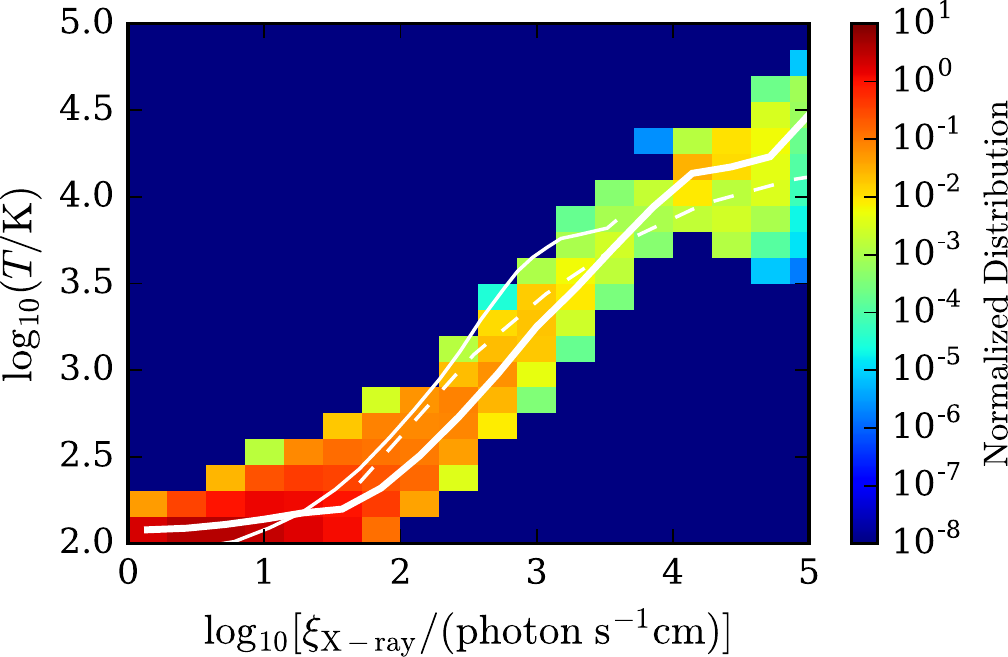}
  \caption{Distribution function of gas in the space spanned
    by
    $\{\log_{10}T\}\times \{\log_{10}\xi_\mathrm{X-ray}\}$,
    measured for the fiducial model. X-ray ionization
    parameter $\xi_\mathrm{X-ray}$ is defined as
    $F_X/n_\chem{H}$, viz. photon number flux density
    divided by the number density of hydrogen nuclei.  Heavy
    white curve indicates the
    $\log_{10}\xi\dash \langle \log_{10}T \rangle$ relation
    based on our colormap histogram (the average of
    $\log_{10}T$ is weighted by the mass of gas), while the
    thin white curves indicate the hydrostatic $\xi\dash T$
    relations adopted in OECA10 (solid: inner-hole disks;
    dashed: primordial disks; note that the definition of
    $\xi$ is different in OECA10, which is converted here to
    the defintion in this paper). }
  \label{fig:phase_xray}
\end{figure}

OECA10 prescribed the gas temperature as a function of the
ionization parameter $F_X/n_H$, where $F_X$ is the local
X-ray flux, taking this relation from the hydrostatic models
of \citet[hereafter EDRCO8]{2008ApJ...688..398E}, who took
$L_X=2\times10^{30}~\erg~\s^{-1}$ (very similar to ours),
though with a slightly harder spectrum ($kT_X=1.5~\keV$).
As shown by Fig.~\ref{fig:phase_xray}, a similar correlation
holds for our fiducial model, though with some spread around
the mean relation.  Recall however that EUV and FUV are also
present in this model, and the former dominates at
$T \gtrsim 10^4~\K$.  At $T\approx 10^2\dash10^3~\K$, the
regime of the intermediate layer, our mean $\xi-T$ relation
falls somewhat below that of OECA10 by
$\sim 0.2\dash0.5~\mathrm{dex}$.  We attribute this
difference partially to the cooling mechanisms. The
hydrostatic models of EDRC08 excluded cooling mechanisms by
molecules and neutral atomic sulfur.
In Table~\ref{table:line-luminosity}, we observe significant
contribution to cooling by \chem{H_2O} and neutral sulphur
in the intermediate layer.

Adiabatic expansion also contributes to the differences
between our $\xi\dash T$ relation and that assumed by
OECA10. Comparing the corresponding panels in
Fig.~\ref{fig:streamline} shows that adiabatic expansion
removes around $3/4$ to $1/3$ (the ratio varies from
location to location) of the internal energy injected by
heating (converting it to kinetic energy);
applying the the hydrostatic $\xi\dash T$ relation
to hydrodynamics assumes that this part of the energy
still contributes to the gas temperature, and hence counts
this part of the energy twice.

Perhaps the dominant difference between our model
and that of EDRC08, however, involves the dust.
Our recipes for thermal accommodation per gas-dust collision
are identical to theirs (see Appendix
\ref{sec:dust-gas-heating}).  Our dust-to-gas ratio
($\sim 10^{-4}$) is smaller than theirs
($6.5\times10^{-4}$).  However, EDRC08 assumed an MRN size
distribution of dust grains \citep{1977ApJ...217..425M}.
The present work assigns a very small radius to all grains
($5~\ang$).  This results in a dust surface area per gas
mass that is $\sim 10$ times larger than in EDRC08, with a
corresponding increase in the rate of cooling by thermal
accomodation.

In support of the arguments here, we have run Model~10
(Table~\ref{table:various-model}),
in which all cooling processes related to
molecules and neutral sulphur are disabled, the abundance
of PAHs is lowered by $\times 10^{-1}$, and X-rays are the
only hard photons (no EUV or FUV).
This model yields a mass-loss
  rate of $\sim 1.1\times 10^{-8}~M_\odot~\yr^{-1}$, fairly
  close to the results in OECA10.

\goodbreak
\section{Conclusions}
\label{sec:conclusion}

In summary, this work combines hydrodynamics with consistent
ray-tracing radiative transfer and thermochemistry to study
the photoevaporation mechanisms of protoplanetary
disks. Irradiated by the FUV, EUV and X-ray radiation from
the central source, the disk develops a three-layer
structure; ordered by increasing latitude, these are a cold
midplane region ($T<10^2\mathrm{K}$), a warm intermediate
layer ($T\sim 10^2\dash10^3\mathrm{K}$) and a hot tenuous
ionized wind ($T\gtrsim 10^4\mathrm{K}$).  The initial
structure of the disk, the abundances and chemistry, and the
amount and size of the dust, broadly follow GH09, but the
time-averaged wind mass-loss rate is somewhat lower,
$\dot{M}_\wind(<10~\au) \approx 3.4\times
10^{-9}~M_\odot~\yr$, and the typical radial velocity is
higher $\sim 30~\kms$, due in part at least to our explicit
multidimensional modeling the hydrodynamics.  Our solutions
never reach a complete steady state, but show fluctuations
$\sim 10$ per cent in $\dot{M}_\wind$, which correlate with what
appears to be an instability in the response of the upper
part of the intermediate layer to irradiation.  (The
fluctuations might be smaller in three-dimensional
simulations due to azimuthal averaging.)  By varying the
various hard-photon luminosities, we find that the most
crucial factor to $\dot{M}_\wind$ is the total heating by
radiation that interacts strongly with molecular and/or
atomic hydrogen, viz. EUV and Lyman-Werner photons. Abundant
soft FUV photons and X-rays also help in launching the wind
by increasing the thickness of the intermediate layer,
causing its upper surface to intercept more of the EUV.  By
comparison with previous work on PPD photoevaporation, we
find that mass-loss estimates are sensitive to the fidelity
of both the hydrodynamics and thermochemistry.  Mass loss is
also sensitive to intrinsically uncertain physical
parameters, notably the dust abundance and size
distribution.  Some molecules, including \chem{OH},
\chem{H_2O} and \chem{CO}, persist in the lower wind at
higher temperatures and intermediate velocities than would
be expected from equilibrium chemistry, as a consequence of
comparable hydrodynamic and (photo-)chemical timescales.

There are several aspects of the problem which we hope to
explore in future work.  Using post-processing of these
hydrodynamic models, now that we have the temperature,
density, and velocity structure, it should be possible to
model molecular and atomic lines, including trace species
not important for heating and cooling, and explicitly
treating the level populations and optical depths of
$\chem{CO}$, etc.  These lines and their profiles will be
used to confront our photoevaporative models with
observations,
e.g. \citet{2004ApJ...603..213C,2008Sci...319.1504C,
  2011A&A...527A.119B, 2013ApJ...770...94B,
  2016ApJ...831..169S}.

Eventually, we plan to add magnetic fields to the problem.
These probably must be present to explain accretion, whether
driven by MRI turbulence or by the wind itself.  The FUV-
and X-ray-heated intermediate layer, which is not hot enough
to escape purely thermally, may well drive a denser, cooler,
and higher-$\dot M$ wind when coupled to the field
\citep{Bai+2016}.

\vspace*{20pt} This work was supported by NASA under grant
NNX10AH37G and by Princeton University's Department of
Astrophysical Sciences.  We thank our colleagues Xuening
Bai, Bruce Draine, Munan Gong, Eve Ostriker, James Owen,
Kengo Tomida, and Zhaohuan Zhu, for discussions and
for detailed comments on a preliminary draft. All errors
are the responsibility of the authors, LW and JG.

%
%
\bibliography{fuv_evap}
\bibliographystyle{apj}

\appendix

\section{Details of Thermochemical processes}
\label{sec:details-chem}

\subsection{FUV induced reactions of \chem{H_2}}
\label{sec:fuv-h2}

When a $13.6~\eV > h\nu > 11.3~\eV$ photon (the
``Lyman-Werner'' band, or LW for short) encounters hydrogen
molecules, this photon can be absrobed by a \chem{H_2}
molecule, and excite the \chem{H_2} molecule into an excited
electronic state. This state can spontaneously decay into
different ro-vibrational states, hence we have to include
the excited \ext{\chem{H_2}} as a representative for excited
molecular hydrogen. Such photo-pumping of \chem{H_2} is also
subject to self-shielding effects.  We follow
\citet{1996ApJ...468..269D} for the shielding factor.

As summarized by TH85, about 10 per cent of the excited
hydrogen molecules result in photo-dissociation; we hereby
simplify this reaction channel by adding a branch to the
main photo-excitation channel, namely,
\begin{equation*}
  \chem{H_2} + h\nu (\mathrm{LW}) \rightarrow
  \begin{cases}
    2\chem{H}\ ,\ & \sim 10\ \text{per cent}\ ; \\
    \ext{\chem{H_2}}\ ,\ & \sim 90\ \text{per cent}\ . \\
  \end{cases}
\end{equation*}
\ext{\chem{H_2}} molecules may also be directly
photo-dissociated; the reaction cross section of TH85 is
adopted.  We take 0.4~eV as the amount of energy deposited
in the gas as heat per FUV dissociation of \chem{H_2}
(\citealt{2016arXiv161009023G}; see also
\citealt{1979ApJS...41..555H}).

At gas densities relevant here, the majority of excited
hydrogen molecules are de-excited by collisions with other
particles, especially \chem{H_2} or H. The de-excitation
rate (with the $v=6$ vibrational state as a proxy for all
excited \ext{\chem{H_2}}), is estimated by (see also TH85),
\begin{equation}
  \label{eq:h2e-deexcite}
  \begin{split}
    k_\de (\chem{H}) & \simeq 1.8 \times
    10^{-13}~\cm^3~\s^{-1}  
    \times \left( \dfrac{T}{\K} \right)^{1/2}
    \exp\left( - \frac{1000~\K}{T} \right) \  ;\\
    k_\de (\chem{H_2}) & \simeq 2.3 \times
    10^{-13}~\cm^3~\s^{-1} 
    \times \left( \frac{T}{\K} \right)^{1/2}
    \exp \left( - \frac{18000~\K}{T + 1200~\K} \right)
    \  .
  \end{split}
\end{equation}
Each collisional de-excitation deposits
$\sim 2.6~\eV$ of heat into the gas (see
also TH85). The rate for spontaneous radiative
de-excitation of \ext{\chem{H_2}} 
is taken to be
$A(\ext{\chem{H_2}}) \simeq 2 \times 10^{-7}~\s^{-1}$
(TH85).

\subsection{FUV induced reactions of \chem{H_2O} and
  \chem{OH}} 
\label{sec:fuv-oh}

Photodissociation of \chem{H_2O} and \chem{OH} is not
drastically affected by self-/cross-shielding due to line
overlap. For the photodissociation cross sections of these
two species as functions of photon energy, we adopt Fig.~1
of AGN14. These reactions also heat the gas. Here we adopt
the estimate in AGN14 that about
$\sim 0.5 (h \nu - E_\diss)$ of heat is deposited into the
gas per reaction, where
$E_\diss (\chem{H_2O}) \simeq 5.13~\eV$, and
$E_\diss (\chem{OH}) \simeq 4.41~\eV$. Photodissociation of
\chem{OH} may result in oxygen atoms in the $^1D$ state,
denoted by \ext{O}, which spontaneously decays to the
$^3P$ state while emitting a photon at $6300~\ang$. Due to
the uncertainty or variability of the FUV spectrum, we adopt
the crude approximation that $\sim 55$ per cent of the
\chem{OH} dissociated results in \ext{O}
\citep{1984Icar...59..305V, 2011A&A...534A..44W}. This seems
not to be significant for hydrodynamics; nevertheless, as
the [O I] $6300~\ang$ radiation is an important diagnostic
of PPDs winds, we expect this to be useful in our incoming
analysis of comparison between simulation results and
observations.

\subsection{Dust and PAH}
\label{sec:dust-pah}

\subsubsection{Dust-assisted \chem{H_2} formation}

The reaction rate of \chem{H_2} formation on dust surface
directly follows \citet{Bai+Goodman2009}, except for the
efficiency of formation, for which we adopt the scheme of
AGN14,
\begin{equation}
  \eta  \sim
  \begin{cases}
    1\ , & T_\dust < 25~\K\ ;\\
    0.6\ , & 25~\K < T_\dust < 80~\K\ ;\\
    0.33\ , & 80~\K < T_\dust < 900~\K\ ;\\
    0\ , & T_\dust > 900~\K\ ;
  \end{cases}
\end{equation}
where $T_\dust$ is the dust temperature, which may be
significantly different from the gas temperature $T$.  A
typical formation rate is
$R\simeq 7.5\times 10^{-17}~\cm^3~\s^{-1}$ at $100~\K$,
which is comparable to the value in photodissociation
regions (PDRs) given the geometric dust cross section
$8\times 10^{-22}~\cm^2$ per hydrogen nucleus \citep[see \S
31.2 in][]{Draine_book}.  The formation of each \chem{H_2}
molecules deposits $\sim 1.5 ~ \eV$ of heat into the gas,
the remaining recombination energy being radiated
(AGN14). We take this effect into account.

\subsubsection{Dust-assisted recombination and photoelectric
  effect}
\label{sec:dust-charge}

Dust-assisted recombination is implemented by
including the following two processes:
\begin{eqnarray*}
  \Gr^+ + e^- \rightarrow \Gr\ ;\quad
  \Gr + e^- \rightarrow  \Gr^-\ .
\end{eqnarray*}
The efficiency of \Gr{} and \pos{\Gr} capturing free
electrons follows the fitting formulae for electrostatic
focusing in \citet{1987ApJ...320..803D} and the sticking
probability evaluated by \citet{2001ApJS..134..263W}. The
following two kinds of reactions close the cycle of
dust-assisted neutralization:
\begin{equation*}
  \Gr^- + \chem{X}^+ \rightarrow X +\G\ ;\quad
  \Gr + X^+ \rightarrow X +\Gr^+\ .
\end{equation*}
Here X represents H, He, C, O, S, Si, or Fe. The rates of
these reactions are evaluated using the same method as in
\citet{Bai+Goodman2009}, using the desorption temperature
summarized by \citet{2006A&A...445..205I}.

Photoelectric reactions of neutral and
negatively charged dust grains are included,
\begin{equation*}
  \Gr+ h \nu \rightarrow  \Gr^+ + e^-\ ;\quad
  \Gr^- + h \nu \rightarrow \Gr+ e^- ~ .
\end{equation*}
The majority of the radiation absorbed is converted to dust
thermal energy (which is in balance with thermal radiation
of dust), but some is carried off by the photoelectrons.
The cross section for photon absorption and photoelectric
yield are evaluated based on the recipes elaborated in
\citet{2001ApJ...554..778L} and
\citet{2001ApJS..134..263W}. The work function is assumed to
be $W\simeq 4.4~\eV$ \citep{2001ApJS..134..263W} for
carbonaceous grains. The energy deposited into the gas per
reaction is estimated by $(h \nu - W)$.  Ideally, our
treatment should involve the valence band ionization
potential \citep{2001ApJS..134..263W}, which differs from
$W$ by $\sim 2~\eV$ for $r_\dust=5~\ang$ particles (this
difference is smaller for larger grains). However, for
simplicity and because our very small grains are proxies for
grains of all sizes, we omit this refinement.

\subsubsection{Dust-gas energy transfer and artificial
  heating term}
\label{sec:dust-gas-heating}

Near the midplane, the gas acquires energy and maintains
temperature through the energy transfer with dust. The
gas-dust energy transfer rate is estimated following
\citet{2001ApJ...557..736G}:
\begin{equation}
  \Lambda_\dust =  \sum_\sp n_\sp
  \left( \dfrac{8 \kb T}{\pi m_\sp} \right)^{1/2}
    \sigma_\dust \alpha \times 2 \kb (T - T_\dust)\ ,
\end{equation}
where the subscripts ``sp'' range over species, $T$ is the
gas temperature, $T_\dust$ the dust temperature, $\sigma_\dust$
is the geometric dust cross section, $\alpha \sim 0.5$ is
the efficiency of gas-dust energy transfer (typically
referred as the {\it accommodation coefficient}). It is
possible that $\Lambda_\dust$ can be negative, indicating a
heating instead of cooling process.

We further assume that the energy-transfer process does not
affect the dust temperature profile. We do not evaluate the
radiative transfer of diffuse infrared radiation iniside the
disk, which should properly determine the temperature of
dust. In order to have a reasonable estimate of $T_\dust$
profile, we assume local equilibrium, and adopt the simplest
dual-temperature profile proposed by
\citet{1997ApJ...490..368C}, using the following equation,
\begin{equation}
  \label{eq:dust-temp}
  0 = \dfrac{\d \epsilon_\dust}{\d t} = \max
  \left[ 4 \sigb T_\ah^4(R) \sigma_\dust q(T_\dust) \ ,\ \sum_{h
      \nu} F_\eff(h \nu) \sigma (h \nu) \right] - 4 \sigb
  T_\dust^4 \sigma_\dust q(T_\dust) \ ,
\end{equation}
where $\sigb$ is the Stefan-Boltzmann constant, $T_\ah$ the
desired artificial heating temperature as a function of $R$
(e.g. \citealt{1997ApJ...490..368C}, figure 4),
$\sigma_\dust$ the geometric cross section of dust, and
$q(T_\dust)$ the Planck-averaged emission efficiency as a
function of black-body radiation field temperature [we
evaluate this value with eq. (24.16) in
\citealt{Draine_book}], $F_\eff(h\nu)$ the local effective
irradiative radiation flux at photon energy bin $h \nu$ (see
eq. \ref{eq:eff-flux}), and $\sigma (h \nu)$ the effective
absorption cross section (see Appendix
\ref{sec:dust-charge}).]

Optical photons ($h\nu \lesssim 4.5~\eV$), which we have not
included in our simulations, should also affect dust
temperature in the regions that those photons penetrate.
Although the optical luminosity is generally $\sim 100$
times greater than all other bands combined, the dust
temperature is rather insensitive to the inclusion of
optical radiation because the thermal emission per grain
$\proptosim T_\dust^6$ (emissivity $q\proptosim T_\dust^2$).
We have conducted a test that includes an optical
photon-energy bin with luminosity $2.34~L_\odot$ (see GH09),
to find that the dust temperature in the intermediate layer
rises by $\sim 40$ per cent, while the gas temperature there is
almost invariant (as gas thermodynamics is dominated by
processes not related to dust in the intermediate layer). As
a result, the mass loss rate is unaffected by optical
photons.

\subsection{Other Molecular and Atomic Cooling Processes}
\label{sec:mol-atom-cooling}

\subsubsection{Molecular ro-vibrational line cooling}
\label{sec:mol-cooling}

Based on \citet{1993ApJ...418..263N}, ro-vibrational cooling
caused by collisionally excited CO, OH, \chem{H_2O} and
\chem{H_2} are evaluated using interpolation tables. All of
those cooling rate calculation schemes require the optical
depth parameter $\tilde{N}_{\mathrm{X}}$ defined in
\citet{1993ApJ...418..263N}, as a measure of escape
probability of photons that remove energy from the gas,
where X is the species interested, and $\tilde N_{\chem{X}}$
is the local $G n (\mathrm{X}) / | \nabla v |$, where $v$ is
the local characteristic velocity, and $G$ is a geometric
factor at the order of 1; note that $\tilde{N}$ has units of
time/volume. Here we use
$\tilde{N}_{\mathrm{X}} \sim n(\mathrm{X})
/(v_{\mathrm{th}}/h)$ to estimate $\tilde{N}$, where $h$ is
the local scale height, and
$v_{\mathrm{th}} \simeq [2 \kb T / m (\mathrm{X})]^{1/2}$ is
the thermal speed of the species X: $n(\mathrm{X}) h$ is a
reasonable estimate of vertical column density integrated
from $z = \infty$.  In the regions where molecular cooling
is important, the magnitude of the vertical gradient in the
flow velocity is $\lesssim 10^{-1}~\kms~\au^{-1}$. This is
comparable to but smaller than $v_{\mathrm{th}}/h$, which is
typically $10^{-1}\dash 10^0~\kms~\au^{-1}$ at the molecular
weights and typical kinetic temperatures of relevant
species.  For simplicity, we use only the thermal speed for
the optical depth parameter.

\subsubsection{Atomic cooling processes}
\label{sec:atom-cooling}

Collisionally excited atoms may decay radiatively, removing
heat from the gas. In this work, we evaluate the cooling
rate of atoms as follows. For each kind of atoms, we assume
that they are in local statistical equilibrium and calculate
the population fraction on the ``upper levels'' of
transitions by taking collisional (de-)excitation, photon or
chemical pumping, and spontaneous decay into account, by
solving detailed balance equations. With the population
number of excited coolants obtained, the cooling rate is
calculated by $\Lambda = \beta n_{\mathrm{coolant}}^* A $,
where $n_{\mathrm{coolant}}^*$ is the number density of the
desired coolant on the excited state, $A$ the Einstein A
coefficient, and $\beta$ the escape probability. According
to previous research work such as
\citet{1981ApJ...250..478K} $\beta$ is a function of
line-center optical depth $\tau_0$,
\begin{equation}
  \beta \simeq
  \begin{cases}
    \left[ \tau_0 \pi^{1 / 2} \left( 1.2 +
        \dfrac{\sqrt{\ln \tau_0}}{1 + \tau_0 / 10^5} \right)
      \right]^{-1}\ , & \tau_0 \gtrsim 1\  ;\\ 
    \dfrac{1 - \exp (- 2 \tau_0)}{2 \tau_0} \ , & \tau_0
    \lesssim 1  \  . 
  \end{cases}
\end{equation}
We estimate $\tau_0$ as a function of local thermal velocity,
vertical column density (estimated by $n_{\mathrm{tot}} h$,
where $h$ is the local scale height), line center
wavelength, and oscillator strength (which can be directly
inferred from $A$), as given by eq. (9.10) of
\citet{Draine_book}. In this work, we include these atomic
coolants: Ly-$\alpha$, [C II] $158~\micron$, [O I]
$63~\micron$, O I $6300~\ang$ (note that the excited state
of oxygen atom for this transition is treated separately;
see Appendix \ref{sec:fuv-oh}), [S I] $25~\micron$, [Si II]
$35~\micron$, [Fe I] $24~\micron$, and [Fe II]
$26~\micron$. We adopt the data in table 4 of TH85 for those
transitions.

%
\end{document}